\documentclass[hidelinks,11pt,a4paper]{article}

\usepackage{jheppub}

\usepackage{mathtools}
\usepackage{bm}
\usepackage{verbatim}
\usepackage{subcaption}
\usepackage{scalerel}
\usepackage{dsfont}
\usepackage{units}
\usepackage{fancyvrb}
\usepackage[toc,page]{appendix}
\usepackage{float}
\usepackage{enumerate}
\usepackage[dvipsnames]{xcolor}
\usepackage[T1]{fontenc}

\usepackage[formats]{listings}
\lstdefineformat{R}{~=\( \sim \),^=\( ^{\wedge} \)}
\newcommand{\mycommentstyle}{\normalfont}
\newcommand{\commentref}[1]{{\mycommentstyle #1}}

\newcommand{\Dx}{\text{d}}
\newcommand{\mt}{m_t}

\newcommand{\ttb}{t\bar{t}}
\newcommand{\CP}{\mathcal{CP}}
\newcommand{\Bhat}{\hat{\beta}_t}

\newcommand{\ct}{c_t}
\newcommand{\ctt}{\tilde{c}_t}
\newcommand{\cti}{c_{t,i}}
\newcommand{\ctj}{c_{t,j}}
\newcommand{\ctti}{\tilde{c}_{t,i}}
\newcommand{\cttj}{\tilde{c}_{t,j}}

\newcommand{\ctone}{c_{t,1}}
\newcommand{\cttwo}{c_{t,2}}
\newcommand{\cttone}{\tilde{c}_{t,1}}
\newcommand{\ctttwo}{\tilde{c}_{t,2}}

\newcommand{\Ztwo}{$\mathbb{Z}_2\ $}
\newcommand{\tree}{\text{tree}}

\newcommand{\MG}{\texttt{MadGraph} }

\newcommand{\ccite}[1]{Ref.\ \cite{#1}}
\newcommand{\ccites}[1]{Refs.\ \cite{#1}}

\usepackage[capitalise]{cleveref} 

\title{Impact of Interference Effects on Higgs-boson Searches in the Di-top Final State at the LHC}

\author[a]{Henning Bahl,}
\author[b]{Romal Kumar,}
\author[b,c]{Georg Weiglein}

\affiliation[a]{Institut f{\"u}r Theoretische Physik,\\Philosophenweg 16, 69120 Heidelberg, Germany}
\affiliation[b]{Deutsches Elektronen-Synchrotron DESY,\\Notkestr.~85, 22607 Hamburg, Germany}
\affiliation[c]{Universit{\"a}t Hamburg,\\Luruper Chaussee 149, 22761 Hamburg, Germany}

\emailAdd{bahl@thphys.uni-heidelberg.de}
\emailAdd{romal.kumar@desy.de}
\emailAdd{georg.weiglein@desy.de}

\preprint{DESY-25-030}

\abstract{The di-top final state is an important search channel for additional Higgs bosons at the LHC. In this channel, large signal--background interference contributions can strongly distort a resonance peak as it would be expected from a pure signal contribution. Moreover, signal--signal interference effects can have a significant impact if more than one additional scalar particle is present. In this work, we perform a comprehensive model-independent analysis of the various interference contributions considering two additional heavy scalars that can mix with each other. We point out the importance of taking into account loop-level mixing between the scalars. A proper treatment of these mixing effects, which has not been previously carried out for the di-top final state, introduces additional relative phases between different parts of the amplitudes entering the interference contributions which we find to have a strong impact on the di-top invariant mass distribution. We study the interference effects both in an idealistic setting as well as taking into account experimental limitations using Monte-Carlo simulations. We demonstrate that the emerging experimental signatures can be unexpected and difficult to interpret. In particular, we point out that an experimental signature manifesting itself as an excess near the $t \bar t$ threshold may actually be caused by new scalar particles with much higher masses. We comment in this context on the recent excess that has been observed by the CMS collaboration near the $t \bar t$ threshold in their searches in the di-top final state.
}

\begin{document}
    \maketitle
    \setlength{\parskip}{3pt}
    \section{Introduction}
The discovery of a Higgs boson with a mass of $125$ GeV~\cite{CMS:2012qbp,ATLAS:2012yve} was a major success for the LHC and motivates the search for additional fundamental scalars that are predicted in models of physics beyond the Standard Model (SM). Beyond the SM (BSM) scalars are for example predicted in extended Higgs sectors such as Two-Higgs Doublet Models (2HDMs) (see~\ccites{Branco:2011iw,Ivanov:2017dad,gunion2018higgs} for reviews on the 2HDM). These models can also induce new sources of $\CP$-violation leading to mixing between BSM scalars.

In many models the BSM scalars have a large coupling to top quarks. Consequently, the di-top final state is a particularly important search channel for BSM scalars with masses above the di-top threshold. The resonant production of a heavy neutral Higgs boson and its subsequent decay to top-quark pairs (the pure signal process) would be expected to manifest itself as a characteristic bump in the invariant mass distribution of the top-quark pair. However, the interference between the resonant production of a heavy Higgs production and the SM QCD background ($gg \rightarrow \ttb$) has a large destructive contribution~\cite{Gaemers:1984sj,Dicus:1994bm,Frederix:2007gi,Craig:2015jba,Jung:2015gta,Bernreuther:2015fts,Carena:2016npr,Hespel:2016qaf,BuarqueFranzosi:2017jrj,Djouadi:2019cbm,RK:2022thesis,Bahl:2024fjb}. This large interference with the SM QCD background leads to a characteristic peak--dip signature in the invariant mass distribution of the top-quark pairs. The signal--background interference patterns depend on the $\CP$ nature of the heavy scalars, their masses, and their decay widths. These interference effects can significantly alter the exclusion limits that are obtained from those searches and leave a considerable parameter region un-excluded that would appear to be ruled out if the interference effects were neglected~\cite{ATLAS:2017snw,CMS:2019pzc}. 

In addition to large (usually destructive) signal--background interference effects, furthermore signal--signal interference contributions can occur if (at least) two neutral heavy scalars contribute to the di-top final state. Signal--signal interference contributions have been studied in~\ccites{Bernreuther:2015fts,Carena:2016npr} for the $t \bar t$ final state, which are mainly focused on the complex 2HDM (c2HDM)~\cite{Ginzburg:2002wt,Khater:2003wq,fontes2018c2hdm}. The phenomenology of signal--signal interference contributions incorporating the relevant loop-level effects and their impact on BSM Higgs searches have furthermore been studied in~\ccites{Fuchs:2014ola,Fuchs:2016swt,Fuchs:2017wkq,Bagnaschi:2018ofa}.

The latest results from experimental searches in the di-top final states can be found in~\ccites{ATLAS:2024itc,ATLAS:2024vxm,CMS:2024ynj}. Interestingly, in 2019 an excess in the invariant mass distribution of the di-top system, making use of spin-correlation information of the produced top quarks that provides information about the $\CP$ nature of a possible BSM state, was found by the CMS collaboration at a mass of about $400$~GeV with a local significance of about $3.5\,\sigma$ based on the partial LHC Run~2 dataset containing $35.9\ \text{fb}^{-1}$ of integrated luminosity~\cite{CMS:2019pzc} (possible interpretations of this excess were discussed e.g.\ in~\ccite{Biekotter:2021qbc}). In the recent CMS result based on the full LHC Run~2 dataset containing $137\ \text{fb}^{-1}$ of integrated luminosity, again exploiting spin-correlation information of the produced top quarks via angular observables (see also~\ccites{Anuar:2024qsz,Arco:2025ydq}), an excess with a local significance of much more than $5\,\sigma$ was found above the perturbative QCD background near the $t \bar t$ threshold~\cite{CMS:2024ynj}. As possible interpretations of this excess in~\ccite{CMS:2024ynj} the cases of a $\CP$-odd Higgs boson $A$ with a mass of about $365$~GeV (having a coupling of $0.75$ times the SM top-Yukawa coupling and a decay width of $2\%$ relative to the mass)\footnote{Within a 2HDM, such an interpretation is in some tension with existing theoretical and experimental bounds, see e.g.\ \ccite{Lu:2024twj}.} and a colour-singlet bound-state type contribution $\eta_t$~\cite{Fadin:1990wx,Fadin:1991zw,Hoang:2000yr,Kiyo:2008bv,Sumino:2010bv,Ju:2020otc,Fuks:2021xje} at a mass of about $343$~GeV were discussed, see also~\ccite{Djouadi:2024lyv}.

In the present work, we study the effect of loop-level mixing between two BSM scalars in the di-top final state. We perform our study in a minimal simplified model (i.e., in a model-independent framework) that involves two additional Higgs bosons with generic $\CP$-mixed couplings. We show that if the scalars mix with each other at the loop level, a rich pattern of signal--background and signal--signal interferences can emerge that can give rise to important modifications of the phenomenology compared to the case of a single new BSM particle. The loop-level mixing is treated via a ``Z-factor'' formalism~\cite{Fuchs:2017wkq,Fuchs:2016swt}. Moreover, we provide all the necessary ingredients to study signal--signal interferences including loop-level mixing in the di-top final using Monte-Carlo event simulations.

The article is organised as follows. In \cref{sec:foundations}, we present our simplified model framework, discuss the various contributions to the di-top final state, and review the Z-factor formalism. In \cref{sec:parton-level-analysis}, as a first step, we perform an analytical parton-level analysis. In \cref{sec:mg-analysis} a Monte-Carlo analysis at the hadronic level is performed, and examples of phenomenologically interesting scenarios are discussed. We provide our conclusions in \cref{sec:conclusions}. \cref{appendix:cross-section-z-fac,appendix:ms-self-energy} contain explicit expressions for the cross-section including the loop-level mixing and the $\overline{\text{MS}}$-renormalised self-energies. \cref{appendix:ufo-details} provides further details on the model file used for Monte-Carlo simulations, while \cref{appendix:sketch-parameters} contains supplementary information on the parameters used. Finally, in \cref{appendix:shape-0-smear} the expected distribution is discussed for the idealised case without experimental smearing of the di-top invariant mass.

	\section{Methodology}
\label{sec:foundations}
\subsection{Simplified model framework}
We perform our study in a minimal simplified model that in addition to the SM particles involves two BSM Higgs bosons which can be $\CP$-mixed states. For our study, we parameterise the top-Yukawa part of the Lagrangian involving the two BSM $\CP$-mixed heavy scalars in the form
\begin{align}
    \label{eq:Lyuk}
	\mathcal{L}_\text{yuk} = -\sum_{j=1}^{2}\dfrac{y_t^\text{SM}}{\sqrt{2}}\,\bar{t}\left(c_{t,j} + i\gamma_5 \tilde{c}_{t,j}\right)t h_{j,\tree} \,,
\end{align}
where $y_t^\text{SM}$ is the SM top-Yukawa coupling, $h_{j,\tree}$ denotes 
the lowest-order mass eigenstates of the heavy scalar particles ($h_{1,\tree}$ and $h_{2,\tree}$), and $t$ and $\bar{t}$ are the top and anti-top quark spinors, respectively. The parameters $c_{t,j}$ and $\tilde{c}_{t,j}$ are the Yukawa-coupling modifiers which rescale the $\CP$-even and $\CP$-odd coupling components of the heavy scalar $h_{j,\tree}$ to the top quark.

In our simplified model approach we do not apply theoretical and experimental constraints on the parameter space. For example, constraints on the free parameters like the tree-level masses and the Yukawa-couplings can be derived from electric dipole measurements or BSM Higgs searches in specific models. Since these constraints depend on details of the models beyond the parametrisation in 
\cref{eq:Lyuk}, their application would require to go beyond the simplified model framework that we use here. The only constraint that we impose for consistency is the requirement that the total decay widths of the heavy scalars have to be at least as large as their partial decay width to top quarks.\footnote{Because of their coupling to top quarks, the BSM scalars can also have a loop-induced decay to gluons and photons. In comparison to the decay mode to top quarks, these channels are, however, suppressed by more than two orders of magnitude. Since, as explained below, our predictions for the Z~factors are based on one-loop contributions, we neglect those loop-induced contributions in the calculation of the minimum decay width.}

\subsection{\texorpdfstring{$\ttb$}{tt} production}
The total scattering amplitude for the process $gg \rightarrow \ttb$ can be written as
\begin{align}
    \mathcal{A} = \mathcal{A}({gg \rightarrow \ttb}) + \sum_{j=1}^{2}\mathcal{A}({gg \rightarrow h_{j,\tree} \rightarrow \ttb}) \,,
\label{eq:amplitude}
\end{align}
where $\mathcal{A}({gg \rightarrow \ttb})$ is the QCD background amplitude and $\sum_{j}\mathcal{A}({gg \rightarrow h_{j,\tree} \rightarrow \ttb})$ is the signal part of the amplitude involving the two BSM Higgs bosons.\footnote{At the same perturbative order, there are additional contributions of the BSM scalars via loop diagrams. As shown in~\ccite{Moretti:2012mq}, these can be sizeable but only have a weak dependence on $m({t\bar t})$. Consequently, their impact is very difficult to distinguish from the background, and we do not further discuss those types of contributions here.} It should be noted that in \cref{eq:amplitude} we do not explicitly account for the (non-resonant) contribution of the SM-like Higgs boson at $125$~GeV. While its interference contribution with the BSM scalars having masses above the $t \bar t$ threshold can safely be neglected, its interference with the QCD background gives rise to a relevant contribution near the $t \bar t$ threshold that should be incorporated in experimental analyses as part of the SM background.

Following~\ccite{Djouadi:2019cbm}, we write the signal part of the amplitude as
\begin{equation}
	\label{expr-signal-amplitude}
	\sum_{j}\mathcal{A}({gg \rightarrow h_{j,\tree} \rightarrow \ttb}) = -\sum_{j} \dfrac{\hat{\Gamma}^{gg h_{j,\tree}}\hat{\Gamma}^{h_{j,\tree}\to t\bar{t}}}{\hat{s} - M^2_{h_{j,\tree}} + iM_{h_{j,\tree}}\Gamma_{h_{j,\tree}}} \,,
\end{equation}
where $\hat{\Gamma}^{gg h_{j,\tree}}$ is the production amplitude for the scalar $h_{j,\tree}$ via gluon--gluon fusion (considering only the top quarks in the loop), and $\hat{\Gamma}^{h_{j,\tree}\to t\bar{t}}$ is the amplitude for its subsequent decay to top and anti-top quarks with four-momenta $k_1$ and $k_2$, respectively. The notation $\hat{\Gamma}$ in the numerator of \cref{expr-signal-amplitude} is used to distinguish the amplitudes of the production and decay processes in the numerator from the decay width $\Gamma$ appearing in the denominator.\footnote{We will use below also the notation of, e.g., $\hat\Sigma$ to indicate a renormalised quantity. The meaning of the notation should be clear from the context, so that no confusion should occur.} The expressions for the production and decay amplitudes are as follows~\cite{Ellis:1975ap,Cahn:1983ip,Spira:1995rr,Plehn:1996wb,Djouadi:2005gi}
\begin{align}
    \label{expr-production-vertex-even-odd} \hat{\Gamma}^{gg h_{j,\tree}} &= \dfrac{\alpha_s}{8\pi v}\hat{s} \left[c_{t,j}A^{H}_{1/2}(\tau_t)\left(g^{\mu\nu} - \frac{p_a^\nu p_b^\mu}{p_a\cdot p_b}\right) + \tilde{c}_{t,j}A^{A}_{1/2}(\tau_t)\frac{p_{a\rho}p_{b\sigma}\epsilon^{\mu\nu\rho\sigma}}{p_a\cdot p_b} \right] \epsilon_{a\mu}\epsilon_{b\nu}\delta_{ab} \nonumber\\
    &\equiv c_{t,j}\hat{\Gamma}_{\text{$\CP$-even}}^{gg h_{j,\tree}} + \tilde{c}_{t,j}\hat{\Gamma}_{\text{$\CP$-odd}}^{gg h_{j,\tree}} \,, \\
    \label{expr-decay-vertex-even-odd} \hat{\Gamma}^{h_{j,\tree}\to\ttb} &= \dfrac{\mt}{v}\bar{u}(k_1)(c_{t,j} + i\gamma_5\tilde{c}_{t,j})v(k_2) \equiv c_{t,j}\hat{\Gamma}_{\text{$\CP$-even}}^{h_{j,\tree}\to\ttb} + \tilde{c}_{t,j}\hat{\Gamma}_{\text{$\CP$-odd}}^{h_{j,\tree}\to\ttb} \,,
\end{align}
where $p_{a,b}$ and $\epsilon_{a,b}$ are the incoming gluon momenta and their polarisation vectors, respectively (with the colour indices $a$ and $b$), $u$ and $v$ are the top-quark spinors, and $v\simeq 246\,\text{GeV}$ is the electroweak vacuum expectation value. It is understood that for the evaluation of the partonic cross-section arising from \cref{eq:amplitude} the appropriate average of the polarisations and colours of the incoming gluons and the sum over the spins of the outgoing top quarks is performed. The loop functions $A_{1/2}^{H,A}$, depending on $\tau_t = \frac{\hat{s}}{4\mt^2}$ with the partonic centre-of-mass energy $\hat s$, are defined in~\ccites{Spira:1995rr,Djouadi:2019cbm}. The partonic differential cross-section for the entire process can then be written as
\begin{align}
	\label{partonic-dcs-1}
	\dfrac{\Dx\hat{\sigma}}{\Dx z} = \dfrac{1}{32 \pi}\dfrac{\Bhat}{\hat{s}}\overline{\left|\mathcal{A}\right|^2} \,,
\end{align}
where $\hat{\sigma}$ is the partonic cross-section and $z =\cos\theta$, with $\theta$ being the scattering angle in the parton-parton centre-of-mass frame between an incoming gluon and an outgoing top quark. $\Bhat \equiv \sqrt{1 - \frac{4\mt^2}{\hat{s}}}$ is the velocity of the top-quark pairs in the centre-of-mass frame.

Considering first the case of a single BSM scalar $h_{j,\tree}$ with coupling modifiers $c_{t,j}$ and $\tilde{c}_{t,j}$, the differential cross-section can be expressed in terms of contributions arising only from the QCD background, the signal of the produced BSM Higgs boson, and the signal--background interference as
\begin{align}
	\label{partonic-dcs}
	\dfrac{\Dx\hat{\sigma}}{\Dx z} = \dfrac{\Dx\hat{\sigma}_\text{B}}{\Dx z} + \dfrac{\Dx\hat{\sigma}_\text{S}}{\Dx z} + \dfrac{\Dx\hat{\sigma}_\text{I}}{\Dx z} \,.
\end{align}
In \cref{partonic-dcs}, the subscript ``B'' denotes the QCD background, ``S'' denotes the signal, and ``I'' denotes the signal--background interference contribution (Sig-Bkg Intf.). The sum of the signal ``S'' and signal--background interference ``I'' cross-section corresponds to the total BSM contribution to the cross-section. The various terms in \cref{partonic-dcs} can be found in~\ccites{Dicus:1994bm,Djouadi:2019cbm}.

The partial decay width of a $\CP$-mixed scalar with the lowest-order mass eigenstate $h_{j,\tree}$ is given by
\begin{align}
	\label{eq:decay-width-minimum}
	\Gamma(h_{j,\tree} \rightarrow t\bar{t}) = 3\dfrac{G_\text{F}\mt^2}{4\sqrt{2}\,\pi}\,\left(c^2_{t,j}\Bhat^3 + \tilde{c}^2_{t,j}\Bhat\right)M_{h_{j,\tree}} \,.
\end{align}
So far, we reviewed the formulae that have already been derived in the literature~\cite{Ellis:1975ap,Cahn:1983ip,Dicus:1994bm,Spira:1995rr,Djouadi:2005gi}.

Before we turn to the cross-sections involving two $\CP$-mixed scalars, it is important to note that for $\CP$-mixed scalars all the cross-terms between $\ct$ and $\ctt$ drop out in the production and decay part in the polarisation-averaged and spin-summed squared amplitude. This means that for a $\CP$-mixed scenario, the $\CP$-even ($\ct$) and $\CP$-odd ($\ctt$) couplings of a $\CP$-mixed scalar appear independently along with their loop-function counterparts (after averaging/summing over the polarisations/spins of the incoming/outgoing particles), i.e.,
\begin{align}
    \label{expr-propto-mixed}
    &\left|\hat{\Gamma}^{ggh_{j,\tree}}\right|^2\biggr|_{\text{$\CP$-mixed}} \propto \left(\ct^2\lvert A^{H}_{1/2}(\tau_\text{t})\rvert^2 + \ctt^2\lvert A^{A}_{1/2}(\tau_\text{t})\rvert^2\right) \,, \\
    &\left\lvert\hat{\Gamma}^{h_{j,\tree}\to\ttb}\right\rvert^2\biggr|_{\text{$\CP$-mixed}} \propto \left(\ct^2 \Bhat^2 + \ctt^2 \right) \,.
\end{align}
We now extend our analysis to include two $\CP$-mixed Higgs bosons. This will be the scenario that we will mainly investigate, first on an analytical level and then using Monte-Carlo simulations. Again employing the definitions of the variables given in \cref{partonic-dcs}, but this time with two $\CP$-mixed Higgs bosons, the expressions for the differential cross-section read (the QCD background term stays the same)
\begin{subequations}
\label{expr-dcs-two-mixed}
\allowdisplaybreaks
\begin{align}
	& \dfrac{\Dx\hat{\sigma}_\text{S}}{\Dx z} = \dfrac{3\alpha^2_\text{s}G^2_\text{F}\mt^2}{8192\pi^3}\hat{s}^2 \left(\dfrac{\left(c_{t,1}^2\Bhat^3 + \tilde{c}^2_{t,1}\Bhat\right)\cdot\left(c_{t,1}^2|A^{H}_{1/2}(\tau_t)|^2 + \tilde{c}_{t,1}^2|A^{A}_{1/2}(\tau_t)|^2\right)}{(\hat{s} - M^2_{h_{1,\tree}})^2 + \Gamma^2_{h_{1,\tree}} M^2_{h_{1,\tree}}}\right. \nonumber \\[3pt]
	\label{expr-dcs-two-mixed-sig} & \hspace{50pt} + \dfrac{\left(c_{t,2}^2\Bhat^3 + \tilde{c}^2_{t,2}\Bhat\right)\cdot\left(c_{t,2}^2|A^{H}_{1/2}(\tau_t)|^2 + \tilde{c}_{t,2}^2|A^{A}_{1/2}(\tau_t)|^2\right)}{(\hat{s} - M^2_{h_{2,\tree}})^2 + \Gamma^2_{h_{2,\tree}} M^2_{h_{2,\tree}}} \\[3pt]
	& \hspace{50pt} \left. +\ 2\times\text{Re}\left[\dfrac{\left(c_{t,1}c_{t,2}\Bhat^3 + \tilde{c}_{t,1}\tilde{c}_{t,2}\Bhat\right) \cdot \Delta(1,2)}{(\hat{s} - M^2_{h_{1,\tree}} + i\Gamma_{h_{1,\tree}} M_{h_{1,\tree}}) \cdot (\hat{s} - M^2_{h_{2,\tree}} - i\Gamma_{h_{2,\tree}} M_{h_{2,\tree}})}\right]\right) \,, \nonumber \\[3pt]
	& \dfrac{\Dx\hat{\sigma}_\text{I}}{\Dx z} = -\dfrac{\alpha^2_\text{s}G_\text{F}\mt^2}{64\sqrt{2}\,\pi}\dfrac{1}{1-\Bhat^2 z^2}\ \scaleobj{1.5}\times \nonumber \\[4pt]
	\label{expr-dcs-two-mixed-intf} & \hspace{30pt} \text{Re}\left[\dfrac{c^2_{t,1}\Bhat^3 A^{H}_{1/2}(\tau_t) + \tilde{c}^2_{t,1}\Bhat A^{A}_{1/2}(\tau_t)}{\hat{s} - M^2_{h_{1,\tree}} + i\Gamma_{h_{1,\tree}} M_{h_{1,\tree}}}\right. \left. + \dfrac{c^2_{t,2}\Bhat^3 A^{H}_{1/2}(\tau_t) + \tilde{c}^2_{t,2}\Bhat A^{A}_{1/2}(\tau_t)}{\hat{s} - M^2_{h_{2,\tree}} + i\Gamma_{h_{2,\tree}} M_{h_{2,\tree}}} \right] \,,
\intertext{where}
	\label{expr-two-scalars-delta} & \Delta(1,2) = \left(c_{t,1} A^{H}_{1/2}(\tau_t)c_{t,2}A^{H,\scaleobj{1.5}*}_{1/2}(\tau_t) + \tilde{c}_{t,1} A^{A}_{1/2}(\tau_t) \tilde{c}_{t,2}A^{A,\scaleobj{1.5}*}_{1/2}(\tau_t)\right) \,, \\
	& \text{and } \tau_t = \dfrac{\hat{s}}{4\mt^2} \,. \nonumber
\end{align}
\end{subequations}
The expressions in~\cref{expr-dcs-two-mixed-sig,expr-dcs-two-mixed-intf} agree with those in the literature in~\ccites{Bernreuther:2015fts,Carena:2016npr}.

One can see that~\cref{expr-dcs-two-mixed-sig} contains, as an important new feature, an interference contribution of the form ($2\times\text{Re}[\cdots]$) in the signal part. We will refer to this contribution as the ``signal--signal'' (or ``sig$1$-sig$2$'') interference term. Therefore, the differential cross-section for the signal part now contains two ``pure''-signal contributions and one signal--signal interference term. The two signal--background interference terms are a straightforward extension of the case with one $\CP$-mixed scalar, i.e., the signal--background interference terms independently occur for the two $\CP$-mixed scalars and hence get added.

\subsection{Treatment of the loop-induced production vertex}
\begin{figure}
    \centering
    \includegraphics[height=0.7\linewidth,width=0.7\linewidth,keepaspectratio]{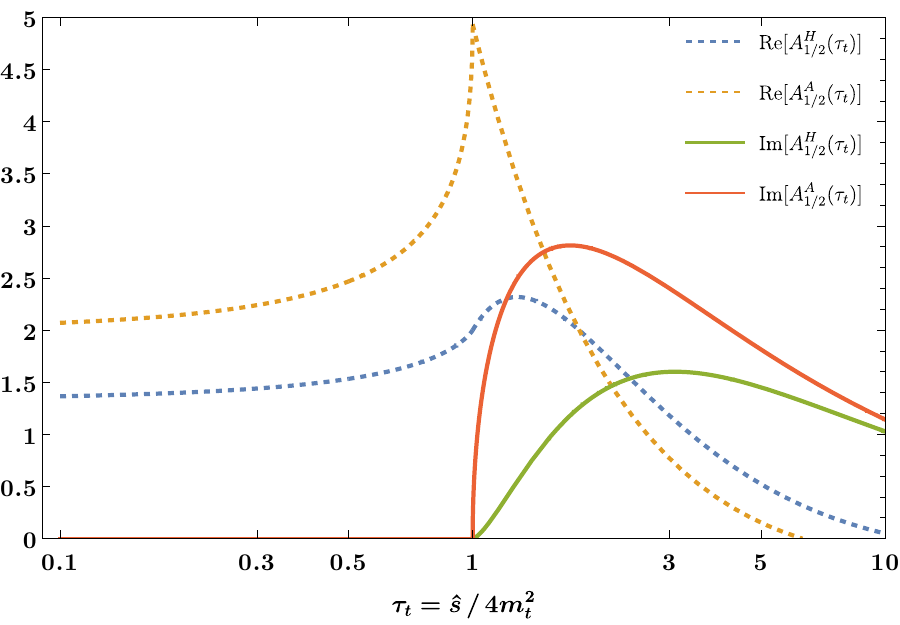}
    \caption{The real and imaginary parts of the top-quark loop function $A^{h_{j,\tree}}_{1/2}$ for the cases of a $\CP$-even ($h_{j,\tree} =~H$) and a $\CP$-odd ($h_{j,\tree} =~A$) state.}
    \label{fig:form-factor}
\end{figure}

For the inclusive on-shell production of the SM-like Higgs boson at $125$~GeV the Higgs--gluon--gluon vertex can be well approximated by a real-valued effective coupling that is obtained in the limit of an infinitely heavy top-quark mass by shrinking the virtual top-quark loop to a point~\cite{Djouadi:1991tka,Dawson:1990zj,Kauffman:1993wv,Demartin:2014fia}. However, for the production of the BSM Higgs bosons considered in this article the limit of an infinitely large top-quark mass would be a poor approximation since the loop-induced Higgs--gluon--gluon production vertex develops a sizeable imaginary part above the threshold where the two top quarks that couple to the Higgs boson can be on-shell. This is illustrated in \cref{fig:form-factor} where the real and imaginary parts of the top-quark loop function are plotted for $\CP$-even and $\CP$-odd scalars. The imaginary part of the loop function becomes non-zero if $\tau \ge 1$, i.e., for the parameter region where the heavy scalar can decay into two on-shell top quarks. The proper incorporation of this imaginary part is crucial for the description of interference effects. In particular, the contribution to the real part in \cref{expr-dcs-two-mixed} arising from the product of the imaginary part in the Breit-Wigner propagator and of the imaginary part from the loop functions is of specific relevance in the resonance region of the two BSM Higgs bosons.
Thus, in order to appropriately simulate the interference effects (and all other contributions as well) we have implemented the full top-triangle loop as a form factor into the Monte-Carlo program that we will use for our numerical analysis below. The implementation details of the full top-triangle loop are described in \cref{appendix:ufo-details}.

\subsection{Higher-order QCD corrections}
\label{sec:k-factors}
All our calculations and simulations are performed at the leading order (LO) in QCD. The production of a Higgs boson (of mass $M_H$) is, however, known to be subject to large higher-order corrections. We incorporate the effect of higher-order corrections at the level of the total cross-section by making use of K~factors for each part of the $t\bar t$ production process.

The K~factors are implemented by scaling the differential cross-section in the invariant mass distribution of the top-quark pairs by the following prescription
\begin{subequations}
\label{expr-hist-scaling-all}
\begin{align}
    \label{expr-hist-scaling-1} \text{Signal$(i)$} &\rightarrow K({\text{sig}(i)})\times\text{Signal$(i)$} \,, \\
    \label{expr-hist-scaling-2} \text{Signal$(i)$--Background interference} &\rightarrow K({\text{sig$(i)$-bkg}})\times \text{Signal--Background interference} \,, \\
    \label{expr-hist-scaling-3} \text{Signal$(i)$--Signal$(j)$ interference} &\rightarrow K({\text{sig$(i)$-sig$(j)$}})\times \text{Signal$(i)$--Signal$(j)$ interference} \,,
\end{align}
where $i$ and $j$ ($j > i$) run over the scalars, and
\begin{align}
    \label{expr-hist-scaling-4} K({\text{sig}(i)}) &= K({\text{production}(i)}) \,, \\
    \label{expr-hist-scaling-5} K({\text{sig$(i)$-bkg}}) &= \sqrt{K({\text{sig}(i)}) \times K({\text{bkg}})} \,, \\
    \label{expr-hist-scaling-6} K({\text{sig$(i)$-sig$(j)$}}) &= \sqrt{K({\text{sig}(i)}) \times K({\text{sig}(j)})} \,.
\end{align}
\end{subequations}
The QCD background ($gg \to \ttb$) K~factor is estimated by comparing the next-to-leading order (NLO) QCD cross-section and the LO QCD cross-section. We take the QCD background K~factor ($K$(bkg)) using the total cross-section for $\ttb$-production (at $13$ TeV centre-of-mass energy at the LHC) at the Next-to-Next-to-Leading Order (NNLO) in QCD to be $1.6$~\cite{Catani:2019hip} (see also~\ccite{Kidonakis:2022hfa}). The K~factor for the signal production process is obtained using the \texttt{HiggsTools} package~\cite{Bahl:2022igd}, which is based on the results of \texttt{SusHi}~\cite{Harlander:2012pb,Harlander:2016hcx} and the recommendations of the LHC Higgs working group~\cite{LHCHiggsCrossSectionWorkingGroup:2016ypw}. We find most of the signal K~factors to be in the range of $2.5$ to $3$, this shows that the NLO contributions are crucial for the signal process and consequently also for the signal--background and the signal--signal processes. 

Finally, we note that a full NLO calculation of the signal--background interference contribution, which is formally a two-loop calculation, was performed in~\ccite{Banfi:2023udd} for the specific case of the extension of the SM by a real singlet. The signal--background interference K~factor that was found for this case in~\ccite{Banfi:2023udd} is slightly smaller than what we consider in this paper.

\subsection{Z~factors and mixing between the scalars}
\label{sec:z-factors}
Besides NLO QCD corrections also electroweak corrections can play an important role. These are particularly relevant for the case of two nearly mass-degenerate states that can mix which each other and can have a strong impact on the mixing pattern. For parametrising this loop-level mixing, we employ the Z-factor formalism as described in~\ccites{Fuchs:2017wkq,Fuchs:2016swt}
(see~\ccites{Dabelstein:1995js,Heinemeyer:2000fa,Heinemeyer:2001iy} for earlier works).

Following the discussion in~\ccite{Fuchs:2016swt}, we consider a system of two tree-level mass eigenstates $h_{1,\tree}$ and $h_{2,\tree}$ that can mix with each other. Neglecting the mixing of those two BSM scalars with SM particles, the inverse propagator matrix of the mixed system will be a $2 \times 2$ matrix, which in general is non-diagonal. Let $(i, j)$ be the indices associated with the tree-level mass eigenstates $(h_{1,\tree},h_{2,\tree})$. Then, the renormalised, one-particle irreducible (1PI) two-point vertex function $\hat{\Gamma}_{ij}(p^2)$ can be written as
\begin{align}
    \hat{\Gamma}_{ij}(p^2) = i\left[(p^2 - m_i^2)\delta_{ij} + \hat{\Sigma}_{ij}(p^2)\right] \,,
\end{align}
where $m_i$ are the tree-level masses and $\hat{\Sigma}_{ij}(p^2)$ are the renormalised self-energies of the two scalars that are expressed in terms of their tree-level mass eigenstates. The associated propagator matrix reads
\begin{align}
    \label{expr-propagator-matrix}
    \bm{\Delta}_{ij} (p^2) = \begin{pmatrix}
        \Delta_{h_{1,\tree}h_{1,\tree}} & \Delta_{h_{1,\tree}h_{2,\tree}} \\
        \Delta_{h_{2,\tree}h_{1,\tree}} & \Delta_{h_{2,\tree}h_{2,\tree}}
        \end{pmatrix}_{ij}
        := - \left(\hat{\bm{\Gamma}}_{ij}(p^2)\right)^{-1} \,.
\end{align}
The matrix inversion yields the individual propagators $\Delta_{ij}(p^2)$ as the ($ij$) elements of the $2\times 2$ matrix $\bm{\Delta}_{ij}(p^2)$. The diagonal propagators $\Delta_{ii}(p^2)$ and off-diagonal propagators $\Delta_{ij}(p^2)$ (with $i \neq j$) can be written as
\begin{align}
    \label{expr:propagator-matrix-element}
    \Delta_{ii}(p^2) & = \dfrac{i\left[D_j(p^2) + \hat{\Sigma}_{jj}(p^2) \right]}{\left[D_i(p^2) + \hat{\Sigma}_{ii}(p^2) \right] \left[D_j(p^2) + \hat{\Sigma}_{jj}(p^2) \right] - \hat{\Sigma}_{ij}^2(p^2)} = \dfrac{i}{p^2 - m_i^2 + \hat{\Sigma}_{ii}^{\text{eff}}(p^2)} \,,\\
    \Delta_{ij}(p^2) & = \dfrac{-i\hat{\Sigma}_{ij}(p^2)}{\left[D_i(p^2) + \hat{\Sigma}_{ii}(p^2) \right] \left[D_j(p^2) + \hat{\Sigma}_{jj}(p^2) \right] - \hat{\Sigma}_{ij}^2(p^2)} \qquad (i\neq j) \,,
\end{align}
where
\begin{align}
    \hat{\Sigma}_{ii}^{\text{eff}}(p^2) = \hat{\Sigma}_{ii}(p^2) - \dfrac{\hat{\Sigma}_{ij}^2(p^2)}{D_j(p^2) + \hat{\Sigma}_{jj}(p^2)} \,, \qquad \text{and} \qquad D_i(p^2) = p^2 - m_i^2 \;.
\end{align}
Finding the poles $\mathcal{M}^2_k$ of the propagator matrix, we obtain the loop-corrected mass eigenstates labelled as $h_1$ and $h_2$, associated to the physical states. The loop-corrected mass eigenstates, which occur as external, on-shell particles (for example, in a decay process) are a mixture of the tree-level mass eigenstates. In the case of $2 \times 2$ mixing, the diagonal ($\Delta_{ii}$) and the off-diagonal ($\Delta_{ij}$) components of the propagator matrix given in \cref{expr-propagator-matrix} each have two complex poles, denoted as $\mathcal{M}_{h_1}^2$ and $\mathcal{M}_{h_2}^2$, in contrast to the case without mixing where the propagator matrix is diagonal and each of the two entries has only a single pole. We order the complex poles according to their real part --- i.e., $\text{Re}(\mathcal{M}_{h_1}^2) \leq \text{Re}(\mathcal{M}_{h_2}^2)$.

The propagator matrix is used to obtain finite wave-function normalisation factors which for external particles ensure the proper normalisation of the S~matrix. Let $(a, b)$ be the indices associated with the loop-corrected mass eigenstates $(h_1, h_2)$. The wave-function normalisation factor for $i$-$j$ mixing on the scalar line for the production or decay part of the amplitude at the pole $\mathcal{M}_a^2$ can be written as a product of the overall normalisation factor $\sqrt{\hat{Z}^{a}_i}$ times the on-shell transition ratio $\hat{Z}^{a}_{ij}$, where
\begin{align}
    \label{expr:os-transition-zfac}
    \hat{Z}^a_i = \left. \dfrac{1}{1 + \dfrac{\partial \hat \Sigma^{\text{eff}}_{ii}(p^2)}{\partial p^2}}\right|_{p^2 = \mathcal{M}_a^2} \,, 
    \qquad \text{and} \qquad
    \hat{Z}^{a}_{ij} = \left. \dfrac{\Delta_{ij}(p^2)}{\Delta_{ii}(p^2)}\right|_{p^2 = \mathcal{M}_a^2} \,.
\end{align}
For the case where $n$ states can mix with each other the assignment between the lowest-order mass eigenstates $i,j,k, \dots$ and the loop-corrected mass eigenstates $a, b, c, \ldots$ is not unique, and it can in fact be shown that all possible assignments of the form $(i,a), (j,b), (k,c), \ldots$ are physically equivalent~\cite{Fuchs:2016swt}. For the case where only two states mix with each other it is convenient to associate the lighter and heavier of the lowest-order mass eigenstates ($h_{1,\tree}, h_{2,\tree}$) with the lighter and heavier one of the loop-corrected mass eigenstates ($h_1, h_2$), respectively. This assignment denoted as $((h_{1,\tree}, h_1), (h_{2,\tree}, h_2))$ implies the evaluation of
\begin{align}
    \label{expr:zfac_assignment_1}
    \hat{Z}^{h_1}_{h_{1,\tree}} \,, \quad
    \hat{Z}^{h_1}_{h_{1,\tree} h_{2,\tree}}
    \quad\text{at}\quad p^2 = \mathcal{M}_{h_1}^2 
\end{align}
and
\begin{align}
    \label{expr:zfac_assignment_2}
    \hat{Z}^{h_2}_{h_{2,\tree}} \,, \quad
    \hat{Z}^{h_2}_{h_{2,\tree} h_{1,\tree}}
    \quad\text{at}\quad p^2 = \mathcal{M}_{h_2}^2 \,.
\end{align}
In this chosen assignment and notation, $\hat{Z}^{h_1}_{h_{1,\tree}h_{1,\tree}} = \hat{Z}^{h_2}_{h_{2,\tree}h_{2,\tree}} =~1 \,.$

The normalisation factors $\sqrt{\hat{Z}^a_i}$ and transition ratios $\hat{Z}^{a}_{ij}$ described in \cref{expr:zfac_assignment_1,expr:zfac_assignment_2} can be arranged into a so-called $\bm{{Z}}$-matrix which is a non-unitary complex matrix.\footnote{This is related to imaginary parts appearing in the propagators of unstable particles.} The concrete $\bm{{Z}}$-matrix for the chosen assignment between the tree-level and loop-corrected mass eigenstates can be written as
\begin{align}
    \label{expr-z-matrix}
    \bm{{Z}} = \begin{pmatrix}
        \sqrt{\hat{Z}^{h_1}_{h_{1,\tree}}}\hat{Z}^{h_1}_{h_{1,\tree}h_{1,\tree}} & \sqrt{\hat{Z}^{h_1}_{h_{1,\tree}}}\hat{Z}^{h_1}_{h_{1,\tree}h_{2,\tree}} \\
        \sqrt{\hat{Z}^{h_2}_{h_{2,\tree}}}\hat{Z}^{h_2}_{h_{2,\tree}h_{1,\tree}} & \sqrt{\hat{Z}^{h_2}_{h_{2,\tree}}}\hat{Z}^{h_2}_{h_{2,\tree}h_{2,\tree}}
    \end{pmatrix} \equiv
    \begin{pmatrix}
        \bm{{Z}}^{h_1}_{h_{1,\tree}} & \bm{{Z}}^{h_1}_{h_{2,\tree}} \\
        \bm{{Z}}^{h_2}_{h_{1,\tree}} & \bm{{Z}}^{h_2}_{h_{2,\tree}} \\
    \end{pmatrix} \,.
\end{align}
One can now make use of the $\bm{{Z}}$-matrix to express the 1PI vertex function $\hat{\Gamma}^{h_1}$ for the production or decay of the loop-corrected mass eigenstate $h_1$ as a linear combination of the 1PI vertex functions for the production or decay of the lowest-order states, $\hat{\Gamma}^{h_{1,\tree}}$ and $\hat{\Gamma}^{h_{2,\tree}}$, as
\begin{align}
    \hat{\Gamma}^{h_1} & = \bm{{Z}}^{h_1}_{h_{1,\tree}}\hat{\Gamma}^{h_{1,\tree}} + \bm{{Z}}^{h_1}_{h_{2,\tree}}\hat{\Gamma}^{h_{2,\tree}} \\
    \label{expr-1pi-z-factor} & = \sqrt{\hat{Z}^{h_1}_{h_{1,\tree}}}\,\left(\hat{\Gamma}^{h_{1,\tree}} + \hat{Z}^{h_1}_{h_{1,\tree}h_{2,\tree}}\hat{\Gamma}^{h_{2,\tree}}\right) \,.
\end{align}
The repeated indices in these expressions are not summed over.

Furthermore, in the simplified model framework used here the 1PI vertex functions of the lowest-order states $(h_{1,\tree}, h_{2,\tree})$ can be expressed in terms of their respective $\CP$-even and $\CP$-odd vertex functions as follows
\begin{align}
    \label{expr:1pi-lowest-order-vertex}
    \hat{\Gamma}^{h_{j,\tree}} = c_{t,j}\hat{\Gamma}_{\text{$\CP$-even}}^{h_{j,\tree}} + \tilde{c}_{t,j}\hat{\Gamma}_{\text{$\CP$-odd}}^{h_{j,\tree}} \,,
\end{align}
see \cref{{expr-production-vertex-even-odd},expr-decay-vertex-even-odd}. Using \cref{expr-z-matrix,expr:1pi-lowest-order-vertex}, the 1PI vertex function $\hat{\Gamma}^{h_1}$ for the production or decay of the loop-corrected mass eigenstate $h_1$ given in \cref{expr-1pi-z-factor} can be expressed in terms of 1PI vertex functions of the lowest-order states and elements of the $\bm{{Z}}$-matrix. Writing the results for $\hat{\Gamma}^{h_1}$ and $\hat{\Gamma}^{h_2}$ conveniently in matrix form yields
\begin{align}
    \label{expr:zfac-vertex-function-even-odd}
    \begin{pmatrix}\nonumber
        \hat{\Gamma}^{h_1} \\[3pt]
        \hat{\Gamma}^{h_2} \\
    \end{pmatrix} &= \bm{{Z}} \cdot \begin{pmatrix}
        c_{t,1}\hat{\Gamma}_{\text{$\CP$-even}}^{h_{1,\tree}} + \tilde{c}_{t,1}\hat{\Gamma}_{\text{$\CP$-odd}}^{h_{1,\tree}} \\[5pt]
        c_{t,2}\hat{\Gamma}_{\text{$\CP$-even}}^{h_{2,\tree}} + \tilde{c}_{t,2}\hat{\Gamma}_{\text{$\CP$-odd}}^{h_{2,\tree}}
    \end{pmatrix} = \\
    &=
    \begin{pmatrix}\left[\bm{{Z}}^{h_1}_{h_{1,\tree}}c_{t,1} + \bm{{Z}}^{h_1}_{h_{2,\tree}}c_{t,2}\right]\hat{\Gamma}_{\text{$\CP$-even}}^{h_{1,\tree}} + \left[\bm{{Z}}^{h_1}_{h_{1,\tree}}\tilde{c}_{t,1} + \bm{{Z}}^{h_1}_{h_{2,\tree}}\tilde{c}_{t,2}\right]\hat{\Gamma}_{\text{$\CP$-odd}}^{h_{1,\tree}} \\[10pt]
        \left[\bm{{Z}}^{h_2}_{h_{1,\tree}}c_{t,1} + \bm{{Z}}^{h_2}_{h_{2,\tree}}c_{t,2}\right]\hat{\Gamma}_{\text{$\CP$-even}}^{h_{2,\tree}} + \left[\bm{{Z}}^{h_2}_{h_{1,\tree}}\tilde{c}_{t,2} + \bm{{Z}}^{h_2}_{h_{2,\tree}}\tilde{c}_{t,2}\right]\hat{\Gamma}_{\text{$\CP$-odd}}^{h_{2,\tree}}
    \end{pmatrix} \,,
\end{align}
where we use the equality of vertex functions for $h_{1,\tree}$ and $h_{2,\tree}$ at the production/decay vertex, i.e., $\hat{\Gamma}^{h_{1,\tree}}_\text{$\CP$-even} = \hat{\Gamma}^{h_{2,\tree}}_\text{$\CP$-even}$ and $\hat{\Gamma}^{h_{1,\tree}}_\text{$\CP$-odd} = \hat{\Gamma}^{h_{2,\tree}}_\text{$\CP$-odd}$ if the vertex functions are evaluated at the same squared momentum (in our case, the square of the centre-of-mass energy).

As shown in~\ccite{Fuchs:2016swt}, the loop-corrected internal propagator of \cref{expr:propagator-matrix-element} taking into account the relevant higher-order contributions is well approximated using Breit-Wigner propagators and the $\bm{{Z}}$-factor matrix,
\begin{align}
    \label{expr-bw-z-factor}
    \Delta_{ii}(p^2) \simeq \sum_{a=h_1,h_2}^{} (\bm{{Z}}^{a}_{i})^2\Delta^{\text{BW}}_a(p^2) \,, \qquad\text{and}\qquad
    \Delta_{ij}(p^2) \simeq \sum_{a=h_1,h_2}^{} \bm{{Z}}^{a}_{i}\Delta^{\text{BW}}_a(p^2)\bm{{Z}}^{a}_{j} \,.
\end{align}
Here, $\Delta^{\text{BW}}_a(p^2)$ is the Breit-Wigner propagator
\begin{align}
    \Delta^{\text{BW}}_a(p^2) = \dfrac{i}{p^2 - \mathcal{M}_a^2} = \dfrac{i}{p^2 - M^2_{a} + i M_a \Gamma_a} \,
\end{align}
with the loop-corrected mass $M_a$ and the total width $\Gamma_a$.

Writing the scalar BSM contributions to the $gg \rightarrow \ttb$ amplitude as
\begin{align}
    \label{eq:prop-mix-amplitude}
    \mathcal{A}_\text{BSM} = \sum_{\substack{i,j=}\substack{h_{1,\tree},\\h_{2,\tree}}} \hat{\Gamma}^{ggi} \Delta_{ij}(p^2) \hat{\Gamma}^{j\to\ttb} \,,
\end{align}
where $\hat{\Gamma}^{ggi}$ and $\hat{\Gamma}^{j\to\ttb}$ are the irreducible vertex functions from the production and decay part of the amplitude, we can approximate it using \cref{expr-bw-z-factor},
\begin{align}
    \label{expr-bw-z-factor-expand}
    \mathcal{A}_\text{BSM} & \simeq \sum_{a=h_1,h_2} \left(\sum_{i=\substack{h_{1,\tree},\\h_{2,\tree}}}\bm{{Z}}^{a}_{i}\hat{\Gamma}^{ggi}\right) \Delta_{a}^{\text{BW}}(p^2) \left(\sum_{j=\substack{h_{1,\tree},\\h_{2,\tree}}}\bm{{Z}}^{a}_{j}\hat{\Gamma}^{j\to\ttb}\right) \,.
\end{align}
Using \cref{expr:1pi-lowest-order-vertex,expr:zfac-vertex-function-even-odd} and (as before) $\hat{\Gamma}^{ggh_{1,\tree}}_{\text{$\CP$-even(odd)}} = \hat{\Gamma}^{ggh_{2,\tree}}_{\text{$\CP$-even(odd)}} \equiv~\hat{\Gamma}^{ggh}_{\text{$\CP$-even(odd)}}$ at the production vertex and $\hat{\Gamma}^{h_{1,\tree}\to\ttb}_{\text{$\CP$-even(odd)}} = \hat{\Gamma}^{h_{2,\tree}\to\ttb}_{\text{$\CP$-even(odd)}} \equiv~\hat{\Gamma}^{h\to\ttb}_{\text{$\CP$-even(odd)}}$ at the decay vertex, \cref{expr-bw-z-factor-expand} can be re-expressed as
\begin{align}
    \label{expr:bw-expand-simple}
    \nonumber & \mathcal{A}_{\text{BSM}} =\\
    \nonumber & \sum_{a=h_1,h_2} \left(\left[\bm{{Z}}^{a}_{h_{1,\tree}}c_{t,1} + \bm{{Z}}^{a}_{h_{2,\tree}}c_{t,2}\right]\hat{\Gamma}^{ggh}_{\text{$\CP$-even}} + \left[\bm{{Z}}^{a}_{h_{1,\tree}}\tilde{c}_{t,1} + \bm{{Z}}^{a}_{h_{2,\tree}}\tilde{c}_{t,2}\right]\hat{\Gamma}^{ggh}_{\text{$\CP$-odd}}\right) \scaleobj{1.2}{\times} \\[2pt]
    \nonumber & \hspace{15pt} \Delta_{a}^{\text{BW}}(p^2)\ \scaleobj{1.2}{\times}  \\
    & \hspace{15pt} \left(\left[\bm{{Z}}^{a}_{h_{1,\tree}}c_{t,1} + \bm{{Z}}^{a}_{h_{2,\tree}}c_{t,2}\right]\hat{\Gamma}^{h\to\ttb}_{\text{$\CP$-even}} + \left[\bm{{Z}}^{a}_{h_{1,\tree}}\tilde{c}_{t,1} + \bm{{Z}}^{a}_{h_{2,\tree}}\tilde{c}_{t,2}\right]\hat{\Gamma}^{h\to\ttb}_{\text{$\CP$-odd}}\right) \,.
\end{align}
It should be noted that at the level of polarisation/spin-averaged squared amplitude the cross-terms between the $\CP$-even and $\CP$-odd parts vanish.

As a result we find that incorporating the effects of loop-level mixing effectively amounts to taking into account the loop-corrected masses arising from the complex poles of the propagators as well as replacing the Yukawa-modifiers by
\begin{subequations}
    \label{expr-ct-ctt-replacements}
    \begin{align}
    \label{expr-ct-ctt-replacements-ct1} c_{t,1} & \rightarrow \bm{Z}^{h_1}_{h_{1,\tree}}c_{t,1} + \bm{Z}^{h_1}_{h_{2,\tree}}c_{t,2} \,, \\
    \label{expr-ct-ctt-replacements-ctt1} \tilde{c}_{t,1} & \rightarrow \bm{Z}^{h_1}_{h_{1,\tree}}\tilde{c}_{t,1} + \bm{Z}^{h_1}_{h_{2,\tree}}\tilde{c}_{t,2} \,, \\
    \label{expr-ct-ctt-replacements-ct2} c_{t,2} & \rightarrow \bm{Z}^{h_2}_{h_{2,\tree}}c_{t,2} + \bm{Z}^{h_2}_{h_{1,\tree}}c_{t,1} \,, \\
    \label{expr-ct-ctt-replacements-ctt2} \tilde{c}_{t,2} & \rightarrow \bm{Z}^{h_2}_{h_{2,\tree}}\tilde{c}_{t,2} + \bm{Z}^{h_2}_{h_{1,\tree}}\tilde{c}_{t,1} \,.
    \end{align}
\end{subequations}
For the case where there is no mixing between the scalars the elements of the $\bm{Z}$-matrix are approximately $\bm{Z}^{h_1}_{h_{1,\tree}} \approx 1 \approx \bm{Z}^{h_2}_{h_{2,\tree}}\,,$ and $\bm{Z}^{h_1}_{h_{2,\tree}} \approx 0 \approx \bm{Z}^{h_2}_{h_{1,\tree}}\,.$ For the general case the evaluation of the $\bm{Z}$-matrix yields the following approximate relations: $\bm{Z}^{h_2}_{h_{2,\tree}} \approx \bm{Z}^{h_1}_{h_{1,\tree}}$ and $\bm{Z}^{h_2}_{h_{1,\tree}} \approx~-\bm{Z}^{h_1}_{h_{2,\tree}}\,.$

It is important to note that the $\bm{Z}$-matrix elements are complex numbers and thus provide an additional source for imaginary parts in the scattering amplitudes besides the ones arising from the loop integrals. As a result, additional phases enter the $gg\to\ttb$ amplitude leading in general to a much richer pattern of interference contributions in comparison to the tree-level case.

Accordingly, in order to take into account the effects of loop-level mixing the analytical equations for the differential cross-section given in \cref{expr-dcs-two-mixed} need to be adapted. For completeness, the explicit expressions incorporating the effects of loop-level mixing are given in \cref{appendix:cross-section-z-fac}.

\subsection{Z-matrix calculation in the simplified model framework}
\label{sec:decay-width-suppression}
We now proceed to the explicit calculation of the elements of the $\bm{Z}$-matrix as defined in \cref{sec:z-factors} in our simplified model framework at the one-loop level. Starting from the lowest-order mass eigenstates ($h_{1,\tree}, h_{2,\tree}$) the loop-corrected mass eigenstates ($h_1, h_2$) are obtained from the complex poles of the propagator matrix. This requires the computation of one-loop self-energies. In this computation, we assume that the top-quark loop is dominant and neglect the contribution from other particles.

While in a complete model one can carry out a renormalisation procedure for the free parameters of the Lagrangian, in our simplified model framework we employ an $\overline{\text{MS}}$ renormalisation for the self-energies. This is not only convenient for simplicity and facilitates the mapping of a complete model to the simplified model framework, it also yields finite expressions for the wave-function normalisation factors that allow the incorporation of numerically important contributions (in particular in the resonance-type region for the mixing between two nearly mass-degenerate states) that are formally of higher order in the perturbative expansion. The expressions for the $\overline{\text{MS}}$-renormalised self-energies are given in \cref{appendix:ms-self-energy}.

The zeros of the determinant of the $2 \times 2$ inverse propagator matrix (see \cref{expr-propagator-matrix}) determine the complex poles at which the Z~factors are evaluated. Explicitly, this reads
\begin{align}
    \label{expr-generic-poles}
    \nonumber & (p^2 - m_{h_{1,\tree}}^2)(p^2 - m_{h_{2,\tree}}^2) \\
    \nonumber \scaleobj{1.1}{+}\  & q(p^2 - m_{h_{1,\tree}}^2)\left[-2\left(\cttwo^2 + \ctttwo^2\right)l_1 + \left(\cttwo^2(p^2 - 4\mt^2) + \ctttwo^2p^2\right)l_2(p^2)\right] \\
    \nonumber \scaleobj{1.1}{+}\  & q(p^2 - m_{h_{2,\tree}}^2)\left[-2\left(\ctone^2 + \cttone^2\right)l_1 + \left(\ctone^2(p^2 - 4\mt^2) + \cttone^2p^2\right)l_2(p^2)\right] \\
    \scaleobj{1.1}{+}\  & q^2\left(\cttwo\cttone - \ctone\ctttwo\right)^2\left(4l_1^2 - 4l_1\left(p^2 - 2\mt^2\right)l_2(p^2) + p^2\left(p^2 - 4\mt^2\right)\left(l_2(p^2)\right)^2\right) = 0 \,,
\end{align}
where we introduced the short-hands $q = \dfrac{3\alpha \mt^2}{8\pi M_W^2 \sin^2\theta_W}$, $l_1 = A_0^\text{fin.}(\mt^2)$, and $l_2(p^2) = B_0^\text{fin.}(p^2, \mt^2, \mt^2) \,.$ Here $A_0$ and $B_0$ are the standard scalar one-loop integrals as defined in~\ccite{Denner:1991kt}, and ``fin.'' denotes their UV-finite parts. In our work, the renormalisation scale (that appears in $\overline{\text{MS}}$-renormalised self-energies) is set to the average of the tree-level masses for our parton-level analysis. For the hadron-level Monte-Carlo study, we set it and the factorisation scale to the average of the loop-level masses. We confirmed that these different choices (made for technical reasons) have a negligible impact on the shown results.

It is interesting in this context to consider the specific case where the last term in \cref{expr-generic-poles} vanishes, i.e., the case where $\left(\cttwo\cttone - \ctone\ctttwo\right)^2 = 0$ or equivalently $t\equiv \nicefrac{\ctone}{\cttone} = \nicefrac{\cttwo}{\ctttwo} \,.$ In this case the determinant of the inverse propagator matrix simplifies to
\begin{align}
    (p^2 - m_{h_{1,\tree}}^2)(p^2 - m_{h_{2,\tree}}^2)
    + q\,\ctttwo^2(p^2 - m_{h_{1,\tree}}^2)h(p^2) + q\,\cttone^2(p^2 - m_{h_{2,\tree}}^2)h(p^2) = 0 \,,
\end{align}
where $h(p^2)\equiv\left[-2\left(1 + t^2\right)l_1 + \left(p^2 + (p^2 - 4\mt^2)t^2\right)l_2(p^2)\right]\,.$ Employing additionally the limit of degenerate tree-level masses, $m_{h_{2,\tree}} \to m_{h_{1,\tree}}$, the determinant of the inverse propagator matrix further simplifies to
\begin{align}
    \label{expr-long-state}
   (p^2 - m_{h_{1,\tree}}^2)\left[p^2 - m_{h_{1,\tree}}^2 + q\left(\cttone^2 + \ctttwo^2\right)h(p^2)\right] = 0 \,.
\end{align}
This implies that in this case one of the solutions for the determinant of the inverse propagator matrix is $p^2 (=\mathcal{M}_{h_1}^2) = m_{h_{1,\tree}}^2\,,$ which is a purely real quantity. Thus, the decay width of one of the loop-corrected states with a pole at $p^2 = m_{h_{1,\tree}}^2$ is zero in this limit. The other solution will, in general, have a non-zero imaginary part.

In fact, in mixing scenarios involving nearly mass-degenerate states, it was indeed found that one of the widths can be suppressed via the so-called ``quantum Zeno effect''~\cite{Sakurai:2022cki,LoChiatto:2024guj}. In the limit where this happens the $\bm{Z}$-matrix is given by
\begin{align}
    \bm{Z} \simeq
    \begin{pmatrix} 
    1/\sqrt{2} & 1/\sqrt{2} \\ -1/\sqrt{2} & 1/\sqrt{2} 
    \end{pmatrix} \,,
\end{align}
corresponding to a maximally mixed scenario.

As a result, for the case where $\cttwo\cttone \approx \ctone\ctttwo$ and the masses of the two states are very close to each other the decay width of one of the loop-level mass eigenstates is suppressed and would go to zero in the limit of the exact coupling relation and mass degeneracy. In our numerical investigation, we will demonstrate that the $\bm{Z}$-matrix plays a crucial role in regulating the resulting sharp peaks in the di-top invariant mass distribution. In \cref{c2hdm-intro}, where we focus on the c2HDM as an example of a potentially realistic model comprising an extended Higgs sector, we will indicate the parameter region where the suppression according to the quantum-Zeno effect can occur.

\subsection{Mapping to the c2HDM}
\label{c2hdm-intro}
As a concrete UV-model that can be mapped to the simplified model framework that we employ in this paper, we consider the complex Two-Higgs Doublet Model (c2HDM). We focus on the c2HDM because it has three neutral scalars, one of which can be identified with the detected SM-like Higgs boson ($h_{\text{125}}$), while the masses for the other two neutral scalars can be above the di-top threshold. In a first step we will relate the parameters of the c2HDM to the Yukawa-coupling modifiers of the simplified model framework. As an application, we will then compare our results to existing results in~\ccite{basler2020di} where the signal--signal interference contributions were not taken into account.

We focus on a c2HDM with a softly-broken \Ztwo symmetry. The \Ztwo symmetry transforms the two Higgs doublets $\Phi_{1,2}$ via
\begin{align}
    \Phi_1 \rightarrow \Phi_1,\qquad \Phi_2 \rightarrow -\Phi_2 \;.
\end{align}
Following closely the discussion in~\ccite{fontes2018c2hdm}, the most general potential is then given by
\begin{align}
    V(\Phi_1,\Phi_2) = &\ m_{11}^2\lvert\Phi_1\rvert^2 + m_{22}^2\lvert\Phi_2\rvert^2 - \left(m_{12}^2\Phi_1^{\dagger}\Phi_2 + \text{h.c.}\right) + \dfrac{\lambda_1}{2}\bigl(\Phi_1^{\dagger}\Phi_1\bigr)^2 + \dfrac{\lambda_2}{2}\bigl(\Phi_2^{\dagger}\Phi_2\bigr)^2 \nonumber \\
    & + \lambda_3\bigl(\Phi_1^{\dagger}\Phi_1\bigr)\bigl(\Phi_2^{\dagger}\Phi_2\bigr) + \lambda_4\bigl(\Phi_1^{\dagger}\Phi_2\bigr)\bigl(\Phi_2^{\dagger}\Phi_1\bigr) + \left[\dfrac{\lambda_5}{2}\bigl(\Phi_1^{\dagger}\Phi_2\bigr)^2 + \text{h.c.} \right] \,.
\end{align}
Due to the hermiticity of the Lagrangian, all couplings are real except for $m_{12}^2$ and $\lambda_5$.

The two doublets $\Phi_i\ (i=1,2)$ are expanded around their real vacuum expectation values (VEVs) $v_1$ and $v_2$, respectively, and the doublets then read
\begin{align}
    \label{expr-c2hdm-doublets}
    \Phi_1 = \begin{pmatrix}
        \phi_1^+ \\
        \frac{1}{\sqrt{2}}(v_1 + \rho_1 + i\eta_1) \\
    \end{pmatrix} \quad , \quad
    \Phi_2 = \begin{pmatrix}
        \phi_2^+ \\
        \frac{1}{\sqrt{2}}(v_2 + \rho_2 + i\eta_2) \\
    \end{pmatrix} \,.
\end{align}
The component fields of the doublets are the charged complex fields $\phi_i^+$ and the real neutral fields $\rho_i$ and $\eta_i$. The Higgs basis $\{\mathcal{H}_1,\mathcal{H}_2\}$ as described in~\ccites{Lavoura:1994fv,Botella:1994cs} is defined by the rotation
\begin{align}
    \begin{pmatrix}
        \mathcal{H}_1 \\
        \mathcal{H}_2
    \end{pmatrix} = R^{T}_{H} \begin{pmatrix}
        \Phi_1 \\
        \Phi_2
    \end{pmatrix} \equiv \begin{pmatrix}
        \cos\,\beta & \sin\,\beta \\
        -\sin\,\beta & \cos\,\beta
    \end{pmatrix} \begin{pmatrix}
        \Phi_1 \\
        \Phi_2
    \end{pmatrix} \,, \qquad \text{with}\qquad  \tan\,\beta = \dfrac{v_2}{v_1} \;,
\end{align}
where
\begin{align}
    \mathcal{H}_1 = \begin{pmatrix}
        G^+ \\
        \frac{1}{\sqrt{2}}(v + H^0 + iG^0) \\
    \end{pmatrix} \quad , \quad
    \mathcal{H}_2 = \begin{pmatrix}
        H^+ \\
        \frac{1}{\sqrt{2}}(R_2 + iI_2) \\
    \end{pmatrix} \,.
\end{align}
The electroweak VEV $v = \sqrt{v_1^2 + v_2^2}$ along with the Goldstone bosons $G^\pm$ and $G^0$ is now rotated to $\mathcal{H}_1$, while the charged Higgs mass-eigenstates $H^\pm$ are part of $\mathcal{H}_2$, and $H^0$, $R_2$, and $I_2$ are real neutral fields. The neutral tree-level Higgs mass-eigenstates $h_{125}$ and $h_{i,\tree}$ $(i=1,2)$ are obtained from the neutral components of the c2HDM basis fields of \cref{expr-c2hdm-doublets} $\rho_1, \rho_2$ (as in the doublets $\Phi_1, \Phi_2$) and $\rho_3 \equiv I_2$ via the rotation
\begin{align}
    \begin{pmatrix}
            h_{125} \\
            h_{1,\tree} \\
            h_{2,\tree}
        \end{pmatrix} = R \begin{pmatrix}
            \rho_1 \\
            \rho_2 \\
            \rho_3
        \end{pmatrix} \,.
\end{align}
After the rotation of the neutral components of the fields in the c2HDM basis, one obtains three neutral tree-level scalars ($h_{\text{125}}, h_{1,\tree}, h_{2,\tree}$) that, in general, are $\CP$-mixed states. Here, we implicitly assume that the states are ordered by their tree-level mass and that the lightest state, $h_{\text{125}}$, corresponds to the detected Higgs boson at $125$~GeV. Additionally, there are two charged scalars $H^\pm$. The $3 \times 3$ mass matrix $\bigl(\mathcal{M}^2\bigr)$ of the neutral scalars is diagonalised via the orthogonal matrix $R$ as described in~\ccite{ElKaffas:2007rq}. That is,
\begin{align}
    \label{eq:c2hdm-mixing-diagonalise}
    R \mathcal{M}^2 R^T = \text{diag}(m_{h_{\text{125}}}^2, m_{h_{1,\tree}}^2, m_{h_{2,\tree}}^2) \,,
\end{align}
with $m_{h_{\text{125}}} \leq m_{h_{1,\tree}} \leq m_{h_{2,\tree}}$. The orthogonal matrix $R$ takes the form
\begin{align}
    R = \begin{pmatrix}
            c_1 c_2 & s_1 c_2 & s_2 \\
            -(c_1 s_2 s_3 + s_1 c_3) & c_1 c_3 - s_1 s_2 s_3 & c_2 s_3 \\
            - c_1 s_2 c_3 + s_1 s_3 & -(c_1 s_3 + s_1 s_2 c_3) & c_2 c_3
    \end{pmatrix} \,,
\end{align}
where $s_i = \sin\alpha_i, c_i = \cos\alpha_i\ (i=1,2,3),$ and $\alpha_i$ are the mixing angles in the orthogonal matrix $R$ that diagonalise the mass matrix ($\mathcal{M}^2$) as illustrated in~\cref{eq:c2hdm-mixing-diagonalise}. Furthermore, $-\pi/2 < \alpha_i \leq \pi/2$\,.

The Yukawa-Lagrangian (in this case comprising of the three neutral scalars of the c2HDM) that describes the couplings of the neutral Higgs bosons to top quarks reads
\begin{align}
    \mathcal{L}_Y = -\sum_{i = 1}^{2} \dfrac{\mt}{v}\,\bar{t}\left(c_{t,i} + i\tilde{c}_{t,i}\gamma_5\right)t \hspace{0.8pt} h_{i,\tree} - \dfrac{\mt}{v}\bar{t}\left(c_{t,h_{\text{125}}} + i\tilde{c}_{t,h_{\text{125}}}\gamma_5\right) t \hspace{0.8pt} h_{\text{125}} \,.
\end{align}
Here the states $h_{i,\tree}$ ($i = 1, 2$) have $\CP$-mixed couplings to the top pairs, while $h_{\text{125}}$ corresponds to the detected Higgs boson at $125$~GeV. Within our simplified model framework the two states $h_{1,\tree}$ and $h_{2,\tree}$ are the two $\CP$-mixed BSM Higgs bosons, 
while the top-Yukawa couplings of $h_{\text{125}}$ do not enter the analysis in our simplified model framework.\footnote{Experimentally it is known that $c_{t,h_{125}}\approx 1$ while $\tilde c_{t,h_{125}}$ so far is only relatively weakly constrained from LHC measurements~\cite{ATLAS:2020ior,CMS:2020cga,Bahl:2020wee,CMS:2022uox,CMS:2022dbt,Bahl:2022yrs,Brod:2022bww,Bahl:2023qwk,ATLAS:2023cbt}.} The relations between the mixing angles ($\alpha_1, \alpha_2, \alpha_3$) and $\tan\beta$ of the c2HDM and the Yukawa-coupling modifiers ($\ctone, \cttone, \cttwo, \ctttwo$) read
\begin{align}
    & \ctone = \dfrac{\cos\alpha_1\cos\alpha_3 - \sin\alpha_1\sin\alpha_2\sin\alpha_3}{\sin\beta}\,, \qquad & \quad & \cttone = -\dfrac{\cos\alpha_2\sin\alpha_3}{\tan\beta} \,, \\
    & \cttwo = -\dfrac{\cos\alpha_1\sin\alpha_3 + \sin\alpha_1\sin\alpha_2\cos\alpha_3}{\sin\beta}\,, \qquad & \quad & \ctttwo = -\dfrac{\cos\alpha_2\cos\alpha_3}{\tan\beta} \,.
\end{align}
Finally, we address the question whether the suppression described above of the decay width of one of the heavy scalars as a consequence of mixing between the heavy scalars is realisable within the c2HDM. As stated in~\cref{sec:decay-width-suppression}, the suppression occurs if the two BSM scalars are nearly mass-degenerate and the relation $\nicefrac{\ctone}{\cttone} \sim \nicefrac{\cttwo}{\ctttwo}$ holds. In the c2HDM, the latter relation corresponds to $\cos\alpha_1 = 0$, which in view of the existing constraints is phenomenologically viable.

	\section{Parton-level analysis}
\label{sec:parton-level-analysis} 
As a next step, we carry out a parton-level analysis using the analytical expressions we presented in \cref{sec:foundations}. In particular, we investigate the background-subtracted BSM cross-section, i.e., the difference between the total cross-section and the contribution arising from the QCD background, as a function of the partonic centre-of-mass energy. The differential signal and signal--background cross-section contributions are integrated with respect to the scattering angle ($\cos\theta = z$),
\begin{align}
    \label{expr-total-bsm-cxn}
    \dfrac{d\sigma^{\text{BSM}}_{\text{Parton}}}{dm(t\bar{t})} = \def\bs{\mkern-15mu} {\scaleobj{1.3}\int}_{\bs -1}^{+1} \left(\dfrac{\Dx\hat{\sigma}_\text{S}}{\Dx z} + \dfrac{\Dx\hat{\sigma}_\text{I}}{\Dx z}\right) \Dx z \,,
\end{align}
where $m(t\bar{t})$ is the di-top invariant mass. In the parton-level analysis that we carry out in the present section we do not fold with a Parton Distribution Function (PDF) but evaluate the cross-section as a function of the partonic centre-of-mass energy.

We do not explicitly discuss the case of a single BSM scalar contributing to the $\ttb$ final state, for which we find very good agreement with the results in the literature~\cite{Gaemers:1984sj,Dicus:1994bm,Bernreuther:2015fts,Carena:2016npr,Djouadi:2019cbm,Frederix:2007gi,Craig:2015jba,Jung:2015gta,Hespel:2016qaf,BuarqueFranzosi:2017jrj}, but focus on the case of two BSM scalars contributing to the $\ttb$ final state. For the case of two mixing BSM scalars in the di-top channel we go beyond the previous results in the literature~\cite{Bernreuther:2015fts,Carena:2016npr} by analysing the implications of loop-level mixing between two $\CP$-mixed scalars. In particular, we will demonstrate the importance of the effects of loop-level mixing for the interpretation of the results from experimental searches.

One can broadly categorise the phenomenology of the different scenarios according to the (tree-level) mass difference between the two heavy $\CP$-mixed scalars,
\begin{enumerate}[(i)]
    \item Large mass separation: the two scalar resonances do not have any significant overlap, $(\Gamma_{h_1} + \Gamma_{h_2}) \ll \lvert M_{h_1} - M_{h_2}\rvert$,
    \item Nearly mass-degenerate: the two scalar resonances significantly overlap $(\Gamma_{h_1} + \Gamma_{h_2}) \gg \lvert M_{h_1} - M_{h_2}\rvert$,
\end{enumerate}
and an intermediate region between the two mentioned cases. In the following, we will describe these scenarios focusing in particular on the nearly mass-degenerate case.

\subsubsection*{Large mass separation --- non-overlapping resonances}
For the case where the tree-level mass separation between the two scalars is larger than the sum of their decay widths, the signal--signal interference contribution is expected to be insignificant compared to all the other contributions. We find that the tree-level and loop-corrected masses are very close to each other in this case. Regarding the Z~factors, we find the off-diagonal elements of the $\bm{Z}$-matrix, $\bm{Z}^{h_1}_{h_{2,\tree}}$ and $\bm{Z}^{h_2}_{h_{1,\tree}}$, to be small. The $\bm{Z}$-matrix is approximately given by an identity ($2 \times 2$) matrix. While for the translation of experimental results into precise model constraints, the shift between the tree-level and loop-level masses should be taken into account, the loop effects do not significantly affect the expected invariant mass distribution (arising from the sum of all the resonance and interference contributions). The invariant mass distribution is characterised by two peak--dip structures that are separated by the mass difference of the two scalars.

\subsubsection*{Nearly mass-degenerate --- significantly overlapping resonances}
We now focus on the scenario where the scalars are nearly mass-degenerate. Nearly mass-degenerate scalars are well-motivated in various extensions of the SM, in which a single mass parameter sets the mass scale of the BSM scalars. In the particular case of $\CP$-violating models, this can lead to mass eigenstates corresponding to highly $\CP$-mixed states where the $\CP$-even and $\CP$-odd Yukawa couplings are comparable in magnitude.

In such scenarios, we find substantial signal--signal interference contributions. Moreover, we find the $\bm{Z}$-matrix to have large off-diagonal elements indicating the relevance of off-diagonal mixing contributions.
\begin{figure}
    \centering
    \includegraphics[width=0.99\textwidth,keepaspectratio]{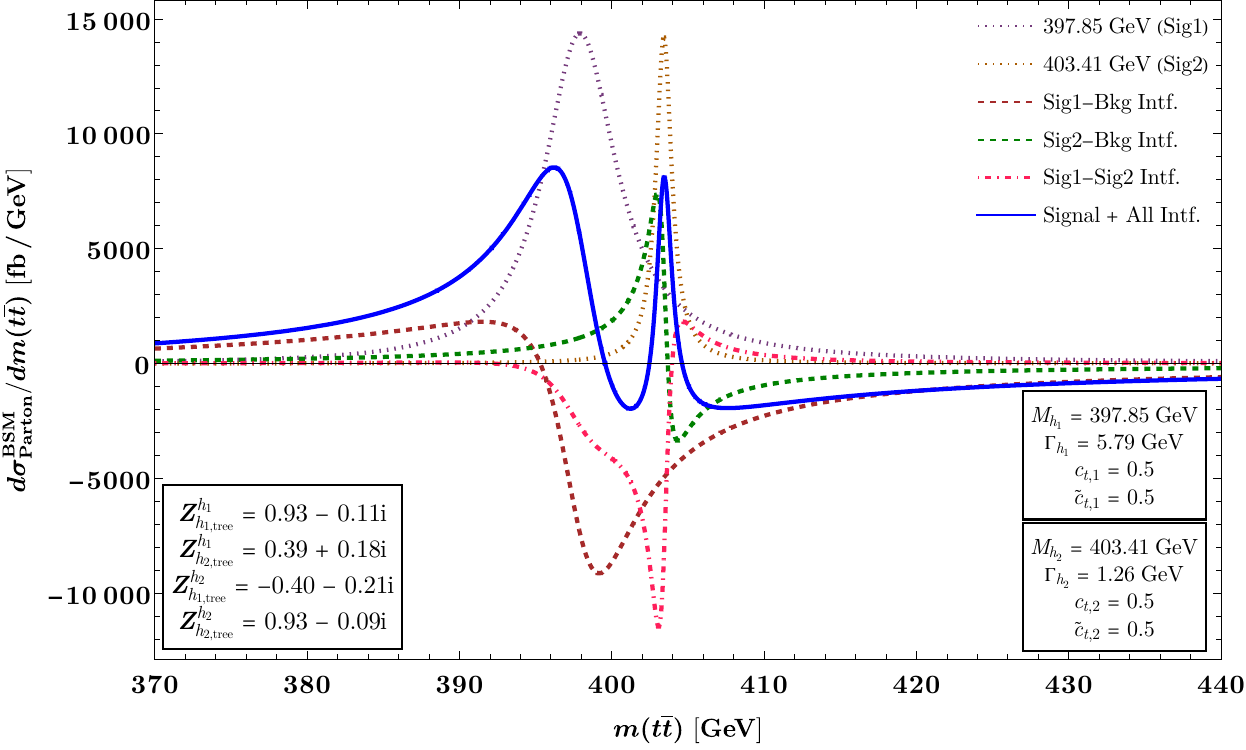}
    \caption{Nearly mass-degenerate scenario: background-subtracted partonic cross-section contributions as a function of the partonic centre-of-mass energy.}
    \label{fig:mass-deg-a}
\end{figure}
In \cref{fig:mass-deg-a}, we show numerical results for such a scenario. In particular, in the considered scenario the mass difference is $\sim 6$~GeV and $c_{t,1} = c_{t,2} = \tilde c_{t,1} = \tilde c_{t,2} = 0.5$. We display the BSM partonic cross-sections versus the di-top invariant mass for all the individual signal and interference contributions and their sum. The colour-code is as follows: the purple dotted curve is the signal resonance, and the brown dashed curve is the signal--background interference for the first scalar. The yellow dotted curve is the signal resonance, and the green dashed curve is the signal--background interference for the second scalar. The red dash-dotted curve denotes the signal--signal interference between the two scalars. The solid blue curve is obtained as the sum of all the different individual contributions. We find a large destructive contribution arising from the signal--signal interference. This feature primarily originates from the fact that the off-diagonal elements of the $\bm{Z}$-matrix are large in magnitude (see also Eq.~(47) of~\ccite{Fuchs:2017wkq}). 
\begin{figure}
    \centering
    \includegraphics[width=0.99\linewidth,keepaspectratio]{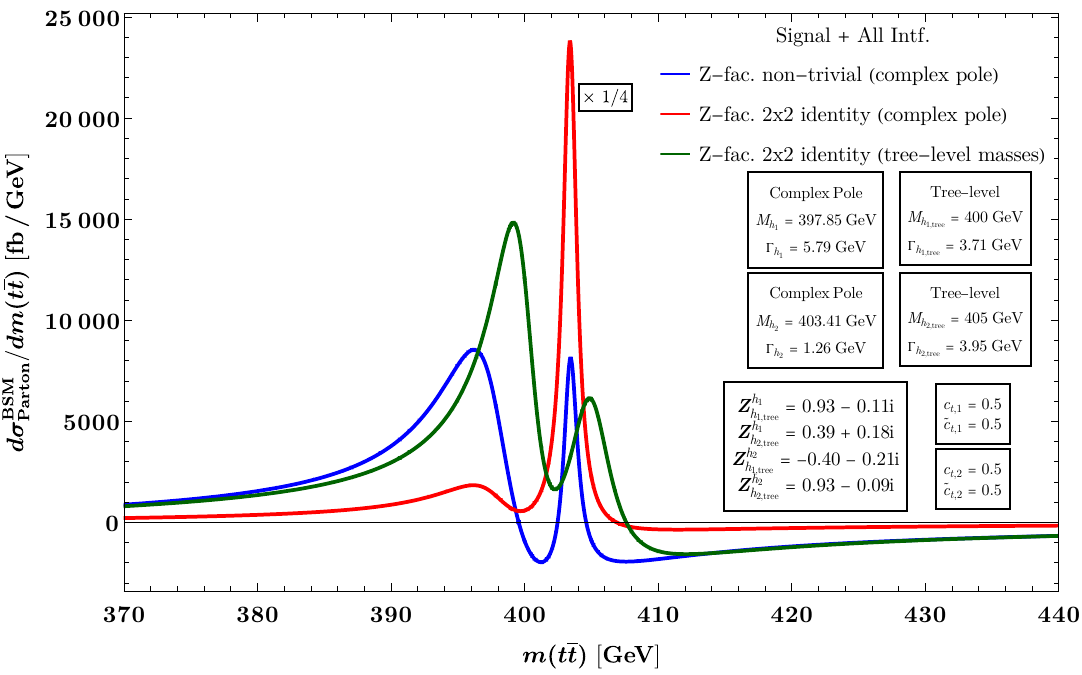}
    \caption{Nearly mass-degenerate scenario: the prediction for the BSM cross-section based on non-trivial Z~factors (blue) is compared with the cases where the $\bm{Z}$-matrix is replaced by a $2 \times 2$ identity matrix (red) and where only the tree-level masses (dark-green) are used.}
    \label{fig:mass-deg-tot-compare}
\end{figure}

In \cref{fig:mass-deg-tot-compare}, we highlight the importance of adequately treating propagator-type mixing between the scalars. The plot contains three different results:
\begin{enumerate}[(i)]
    \item ``Z-fac. non-trivial (complex pole)'' (blue): The complex poles and the $\bm{Z}$-matrix are used. This result corresponds to the most precise prediction for the considered scenario.
    \item ``Z-fac. $2 \times 2$ identity (complex pole)'' (red): The complex poles are used, but the $\bm{Z}$-matrix is set to an identity $2 \times 2$ matrix ($\mathds{1}_2$).
    \item ``Z-fac. $2 \times 2$ identity (tree-level masses)'' (green): Only the tree-level masses and the minimum decay width as obtained from \cref{eq:decay-width-minimum} are considered with the $\bm{Z}$-matrix set to the $\mathds{1}_2$ matrix. This corresponds to a tree-level approximation.
\end{enumerate}

The scenario investigated in \cref{fig:mass-deg-tot-compare} is such that it leads to a suppression of the decay width of $h_2$ as a consequence of loop-level mixing according to the quantum-Zeno effect, as explained in \cref{sec:decay-width-suppression}. If the effects of loop-level mixing are not properly incorporated into the $\bm{Z}$-matrix but instead an identity $2 \times 2$ matrix is used, the suppression of the decay width of $h_2$ is seen to give rise to a huge peak in the invariant mass distribution. On the other hand, for the result based on the properly calculated $\bm{Z}$-matrix the peak is regulated leading to a curve which is more similar to the tree-level result but still significantly deviates from it. The difference to the tree-level result is partly a consequence of the additional phases in the $\bm{Z}$-matrix elements. The displayed results clearly show the importance of accurately taking into account loop-level mixing for preventing the appearance of artificially large resonance peaks in the invariant mass distribution.
\begin{figure}
    \begin{subfigure}{0.49\textwidth}
        \centering
        \includegraphics[width=0.99\textwidth,keepaspectratio]{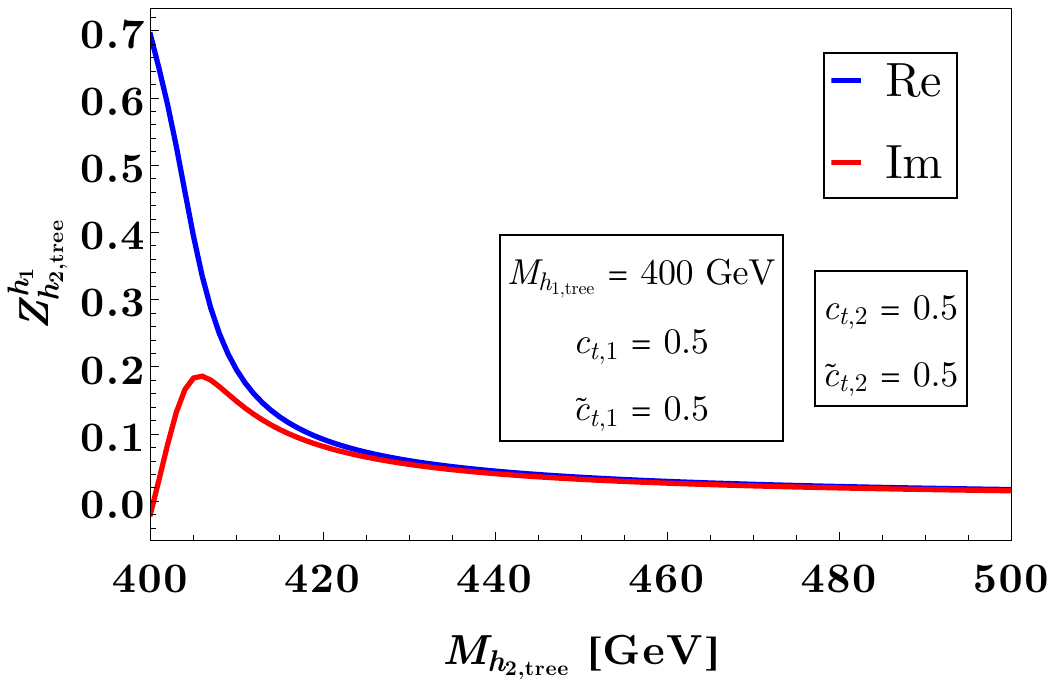}
        \caption{$\bm{Z}^{h_1}_{h_{2,\tree}}$-element of the $\bm{Z}$-matrix}
        \label{fig:zfac-400-scan-z12}
    \end{subfigure}%
    \begin{subfigure}{0.49\textwidth}
        \centering
        \includegraphics[width=0.99\textwidth,keepaspectratio]{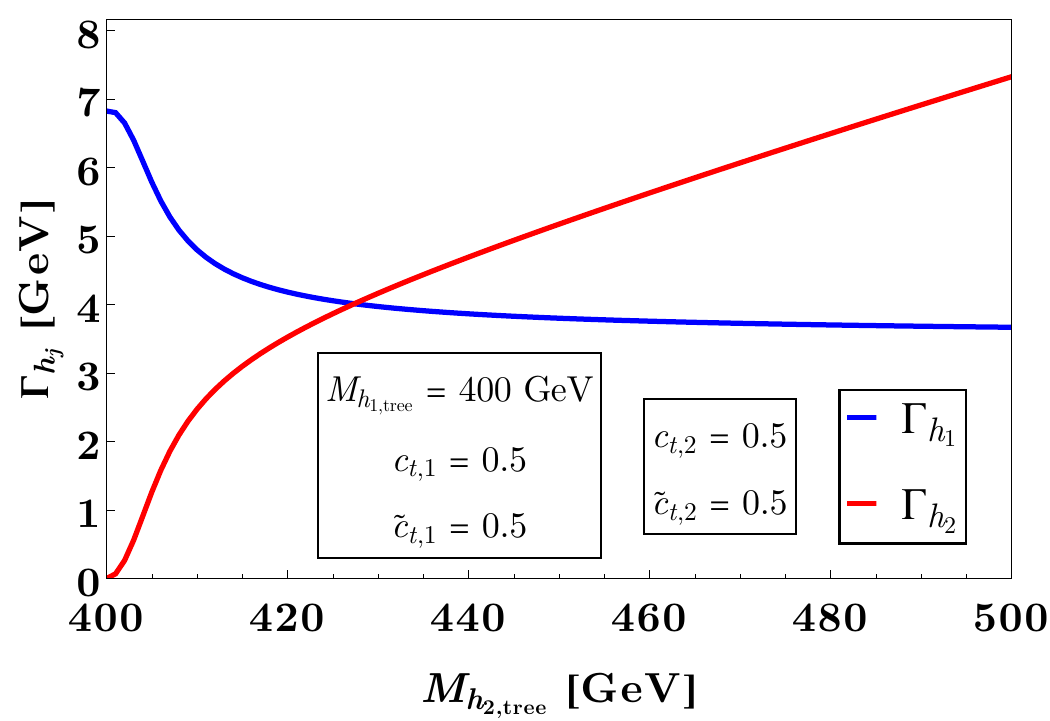}
        \caption{Decay widths of the loop-corrected states}
        \label{fig:zfac-400-scan-width}
    \end{subfigure}
    \caption{(a): One of the off-diagonal elements of the $\bm{Z}$-matrix, $\bm{Z}^{h_1}_{h_{2,\tree}}$, as function of the tree-level mass $M_{h_{2,\tree}}$ for a scenario with $M_{h_{1,\tree}} = 400$ GeV where both scalars have a $\CP$-even and $\CP$-odd Yukawa-coupling of $0.5$\,. (b): Same as (a) but the decay widths $\Gamma_{h_j}$ of the two loop-corrected states into top-quark pairs are shown.}
    \label{fig:zfac-400-scan}
\end{figure}

To conclude our parton-level analysis, we study in \cref{fig:zfac-400-scan} how the elements of the $\bm{Z}$-matrix evolve as one of the tree-level masses is varied while keeping all other input parameters fixed (\cref{fig:zfac-400-scan-z12}), and similarly how the decay widths evolve in the same scan (\cref{fig:zfac-400-scan-width}). It is evident from \cref{fig:zfac-400-scan-z12} that the $\bm{Z}$-matrix is indeed approaching the $\bm{Z}$-matrix associated with a maximally mixed scenario as described in \cref{sec:decay-width-suppression} if the two masses get close to each other. The suppression of one of the decay widths --- demonstrating the quantum-Zeno effect --- is clearly seen in \cref{fig:zfac-400-scan-width} as one of the tree-level masses ($M_{h_{2,\tree}}$) approaches the other tree-level mass, which is kept fixed here at $M_{h_{1,\tree}} = 400$~GeV.

	\section{Monte-Carlo analysis at the hadronic level}
\label{sec:mg-analysis}
After studying the interference effects at the partonic level, we now turn to the hadronic level. In particular, we implement our simplified model framework into the Monte-Carlo code \MG (version 3.4.0)~\cite{Alwall:2014hca}, fold the differential cross-section with the gluon PDFs, and perform an approximate incorporation of detector effects. For the \textit{Universal FeynRules Output} (UFO) input model, we use a custom model file that extends the SM by two scalars with the Yukawa couplings defined in \cref{eq:Lyuk}. The model file includes support to input arbitrary (complex) Z~factors. The scalar--gluon--gluon interaction is implemented via a form factor taking into account the full dependence on the momentum and the mass of the top quark in the loop. We validated our model by comparing it to the analytical results presented in \cref{sec:parton-level-analysis} finding good agreement. For more details, see \cref{appendix:ufo-details}.

We estimate the experimental sensitivity at current or future LHC runs following the procedure outlined in~\ccite{Anuar:2024qsz}, which is based on the CMS analysis presented in~\ccite{CMS:2019pzc}. While the experimental analyses consider several final states, we concentrate here on the fully leptonic final state. To account for the top-quark decays, we multiply the di-top cross-section with the leptonic top-quark branching ratios. Furthermore,~\ccite{CMS:2019pzc} performs its analysis in different regions of the spin correlation variable $c_\text{hel}$. Focusing on the most sensitive region ($0.6 < c_\text{hel} < 1$) as in~\ccite{Anuar:2024qsz}, we approximately account for the selection by multiplying with the selection efficiency, which we estimated to be $\sim 0.18$. To estimate the experimental sensitivity, we compare our differential $m({\ttb})$ distributions (obtained by treating the top quarks as stable) to the post-fit uncertainty band for the di-lepton final state from~\ccites{CMS:2019pzc,Anuar:2024qsz}. In particular, we account for a statistical uncertainty in the background, which is obtained by taking the square root of the expected SM $\ttb$ background for a luminosity of $300\ \text{fb}^{-1}$ at the LHC (corresponding to the prospective integrated luminosity after Run~3 of the LHC) with $13$ TeV centre-of-mass energy.\footnote{The expected number of events for a centre-of-mass of $13.6$ TeV changes slightly compared to the case for $13$ TeV centre-of-mass energy. Changes in the shape of the distribution are negligible.} If the predicted BSM contribution exceeds this statistical uncertainty band, we expect the LHC to become sensitive to the experimental signature of such a BSM contribution (for instance, during the partial or full Run~3 data analysis or during the High-Luminosity phase of LHC). Finally, to account for the limited experimental mass resolution, we apply a Gaussian smearing of $15\%$ to the $m(\ttb)$ values of the generated Monte-Carlo events. For the binning of both the experimental uncertainty band and the BSM contribution, we orient ourself on the current experimental resolution choosing a bin width of $\sim 30$~GeV.

\subsection{Overview of different interference patterns}
\label{sec:overview-15-smear}
\begin{figure}
    \centering
    \includegraphics[width=0.99\textwidth,keepaspectratio]{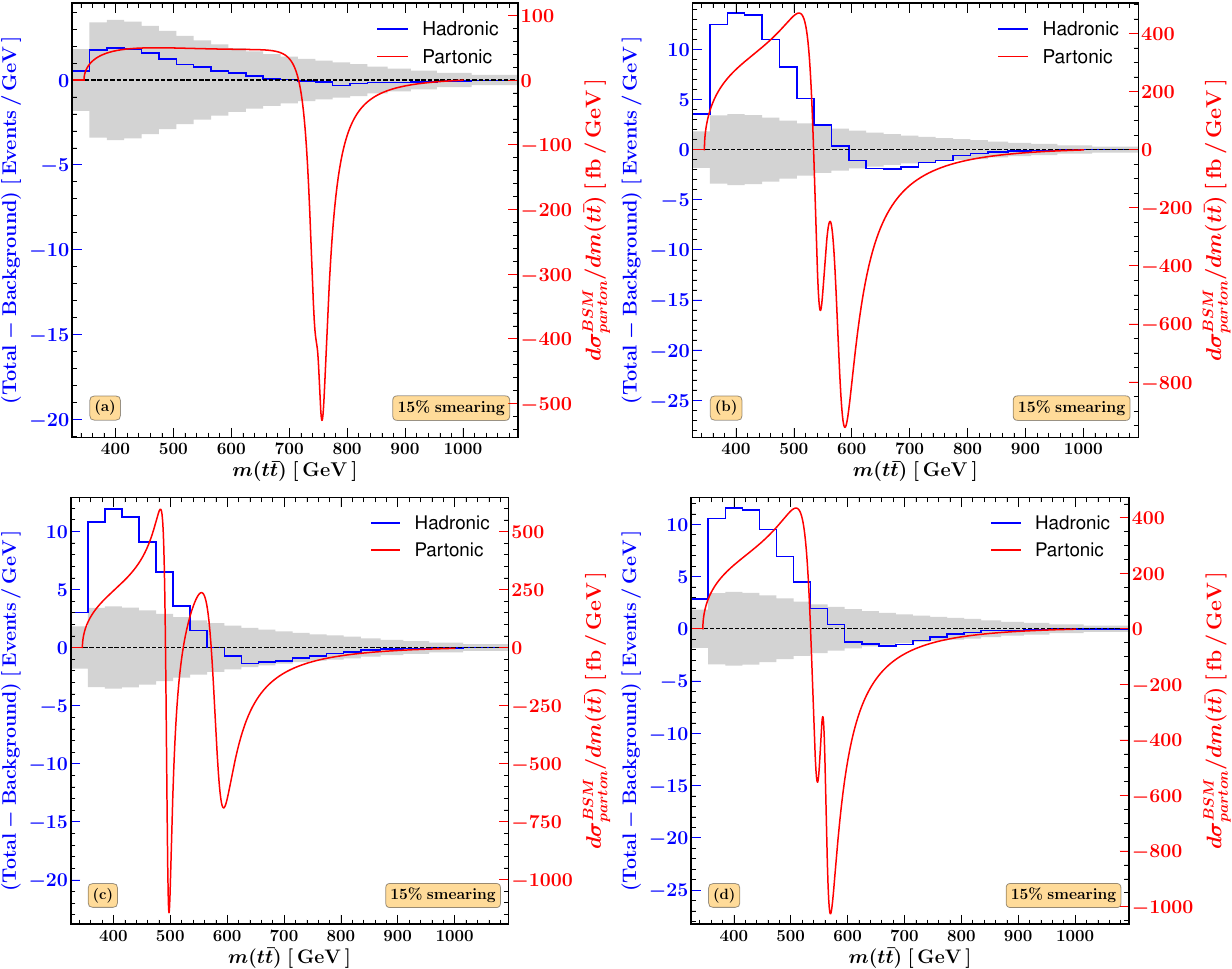}
    \caption{(a)--(d) Background-subtracted $m(\ttb)$ distributions resulting from the sum of all resonance and interference contributions. The plots overlay the partonic-level (shown in red) and the hadronic-level (shown in blue) expectations in the $m(\ttb)$ distribution. The grey band denotes the statistical uncertainty on the SM $\ttb$ background at the hadronic level.}
    \label{fig:lineshape-sketch-15}
\end{figure}
In order to provide an overview of the (approximate) impact of the folding with the gluon PDFs as well as the limited experimental resolution on the $m(\ttb)$ distributions, we present in \cref{fig:lineshape-sketch-15} the background-subtracted $m(\ttb)$ distributions both at the partonic (red curves) and hadronic level (blue curves) for four illustrative scenarios. For the PDFs, we use the NNPDF2.3 set~\cite{Ball:2012cx} accessed using LHAPDF~\cite{Buckley:2014ana}. The grey band indicates the statistical uncertainty of the SM QCD background at the hadronic level. The input parameters used for obtaining the various partonic-level and hadronic-level curves in \cref{fig:lineshape-sketch-15} are listed in \cref{appendix:sketch-parameters}. For comparison, the same plots as in in \cref{fig:lineshape-sketch-15} are shown for the case of $0\%$ smearing, thus displaying only the impact of folding with the gluon PDFs, in \cref{fig:lineshape-sketch-0} in \cref{appendix:shape-0-smear}.

We observe that the Gaussian smearing that is applied in order to account for the limited experimental resolution washes out many features of the parton-level distributions. Consequently, distinctive features of BSM contributions at the parton level BSM can turn out to be very difficult to detect at the hadron level. The folding with the gluon PDFs results in an enhancement of the distributions for low $m(\ttb)$. This effect can generate broad peaks or even a plateau-like behaviour below the actual masses of the BSM particles.

\subsection{Exploratory benchmark scenarios at the LHC}
\label{sec:benchmark-scenarios}
\begin{figure}
    \centering
    \includegraphics[height=0.9\linewidth,width=0.9\linewidth,keepaspectratio]{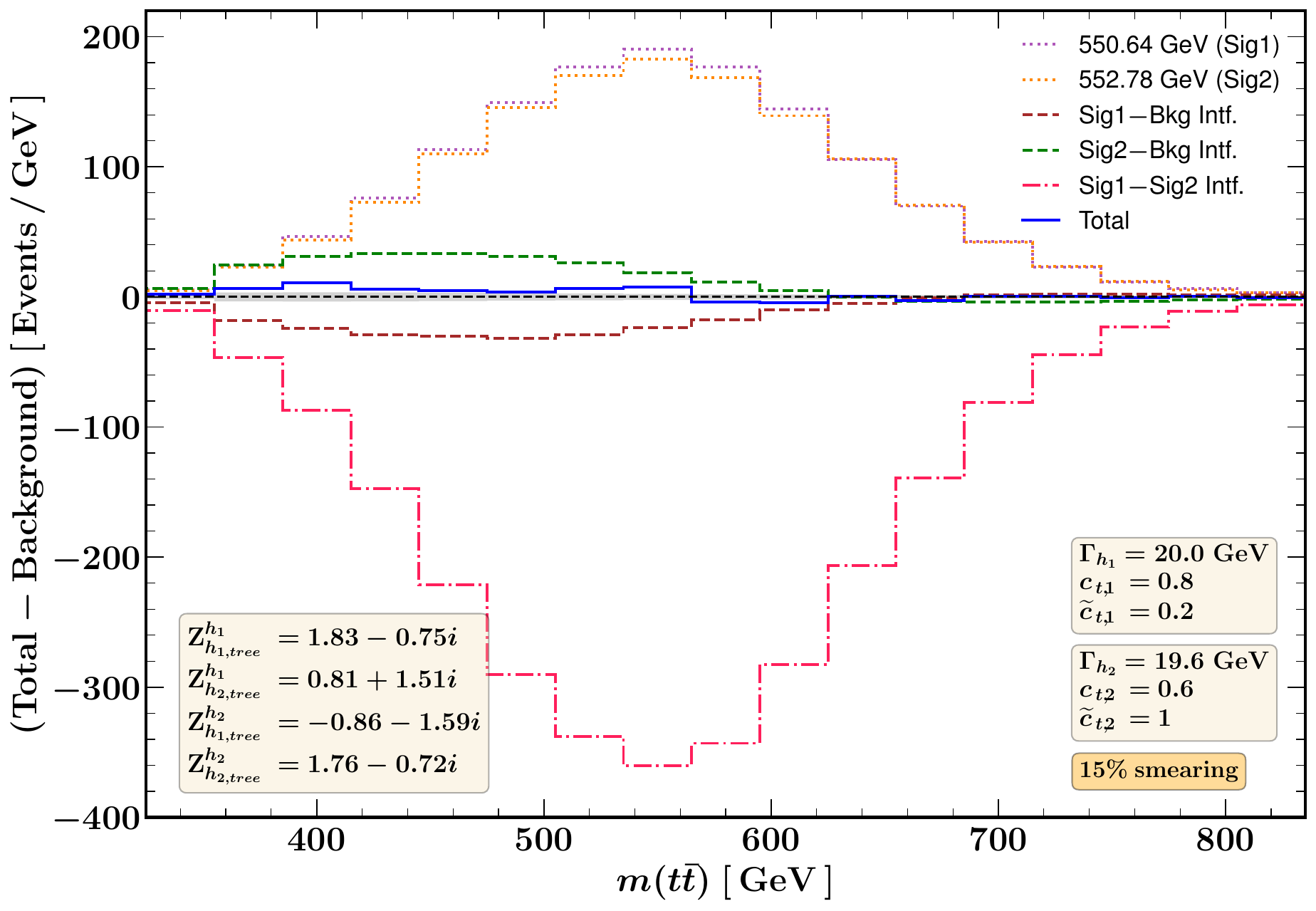}
    \caption{Scenario involving large cancellations: contributions to the background-subtracted $m(\ttb)$ distribution at the hadronic level. The tree-level masses are $550$ GeV and $571$ GeV. All the other parameters are indicated in the plot.}
    \label{fig:long-plateau-scenario}
\end{figure}

\begin{figure}
    \centering
    \includegraphics[height=0.9\linewidth,width=0.9\linewidth,keepaspectratio]{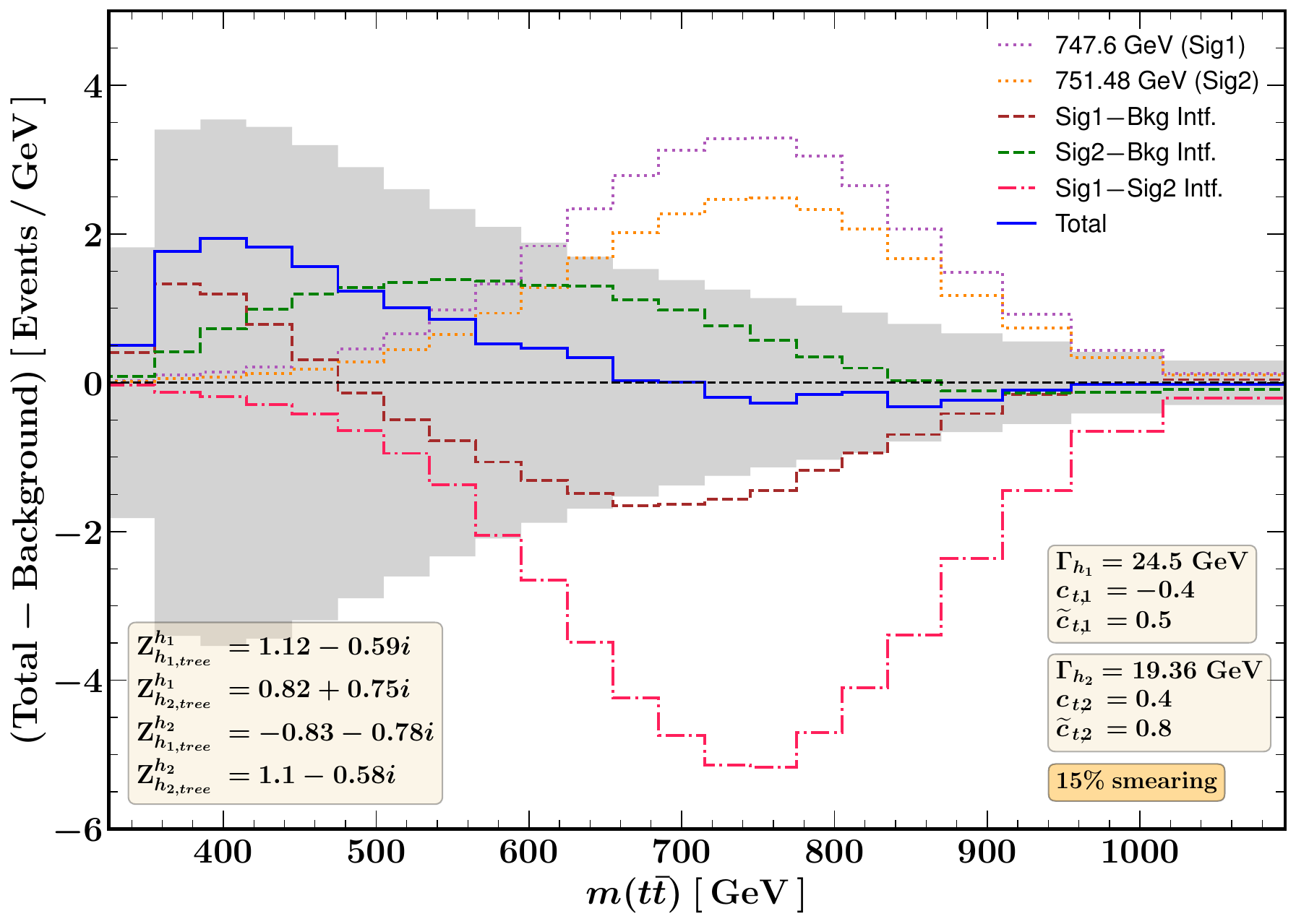}
    \caption{Scenario giving rise to a distribution that lies within the experimental uncertainty band (grey band): contributions to the background-subtracted $m(\ttb)$ distribution at the hadronic level. The tree-level masses are $750$ GeV and $766$ GeV. All the other parameters are indicated in the plot.}
    \label{fig:nightmare-scenario}
\end{figure}

\begin{figure}
    \centering
    \includegraphics[height=0.9\linewidth,width=0.9\linewidth,keepaspectratio]{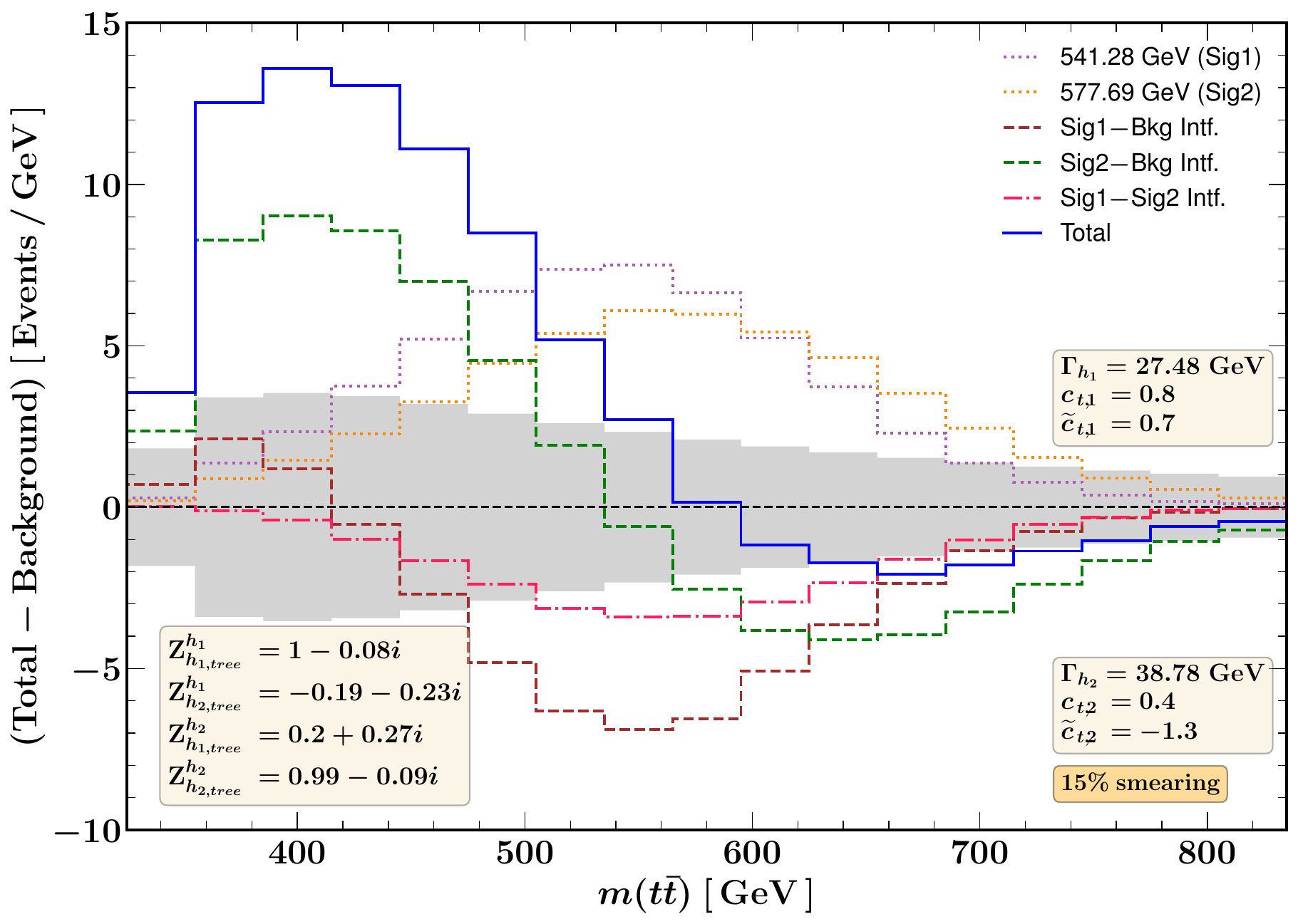}
    \caption{Close-in-mass scenario: contributions to the background-subtracted $m(\ttb)$ distribution at the hadronic level. The tree-level masses are $550$ GeV and $600$ GeV. All the other parameters are indicated in the plot.}
    \label{fig:low-mass-resonance-scenario}
\end{figure}

\begin{figure}
    \centering
    \includegraphics[height=0.9\linewidth,width=0.9\linewidth,keepaspectratio]{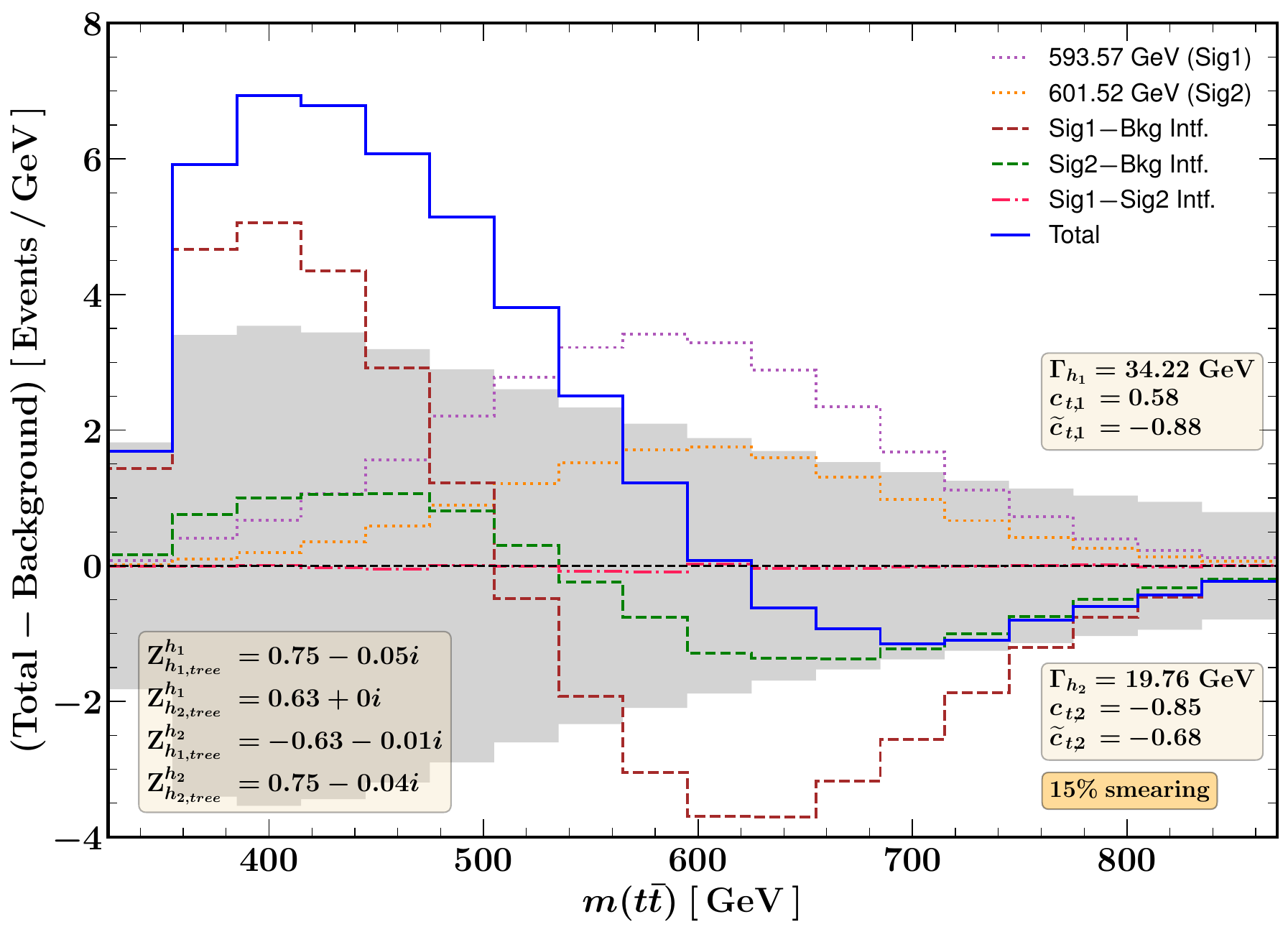}
    \caption{c2HDM scenario: contributions to the background-subtracted $m(\ttb)$ distribution at the hadronic level. The tree-level masses are $608.6$ GeV and $609.4$ GeV. All the other parameters are indicated in the plot.}
    \label{fig:full-c2hdm-bp3-scenario}
\end{figure}

While for the overview shown in \cref{fig:lineshape-sketch-15} only the total result was displayed for each scenario, we now investigate in detail for various benchmark scenarios the interplay between the different BSM contributions and show the results in \cref{fig:long-plateau-scenario,fig:nightmare-scenario,fig:low-mass-resonance-scenario,fig:full-c2hdm-bp3-scenario}. The colour-coding in those figures is as follows: the purple (orange) dotted curve denotes the contribution of the pure signal resonance for the loop-level mass eigenstate $h_1$ ($h_2$), where the loop-level mass of the resonance is indicated in the plot legend with Sig$1$ (Sig$2$) in the parenthesis. The brown (green) dashed curve shows the signal--background interference contribution of the scalar $h_1$ ($h_2$) with the QCD background, denoted as Sig$1$--Bkg Intf.\ (Sig$2$--Bkg Intf.). The red dash-dotted curve corresponds to the signal--signal interference contribution between the two scalars and is marked as Sig$1$--Sig$2$ Intf.\ in the plot legend. The curves have been appropriately scaled by the NLO K~factors\footnote{While we separately calculate the K~factors for each considered scenario, it is useful to note that we find values of about $2.5$ for the signal K~factors, and hence the signal--signal interference K~factor is also about~$2.5$. For the signal--background interference K~factors we find values of around~$2$ (given a QCD background, $gg \to \ttb$, K~factor of~$1.6$).} at the histogram level as described in \cref{sec:k-factors}. The solid blue line displays the sum of all the signal and interference contributions.

We start with a scenario in \cref{fig:long-plateau-scenario} with two $\CP$-mixed scalars with tree-level masses of $550$ and $571$ GeV. We see that this choice of parameters leads to loop-level mass eigenstates that are nearly mass-degenerate. This gives rise to a large resonant-type mixing between the two scalars with large real and imaginary pieces in the elements of the $\bm{Z}$-matrix, in particular for the off-diagonal terms. Due to large absolute values of the elements of the $\bm{Z}$-matrix, all the individual contributions are also enhanced. However, the sum of all contributions involves large cancellations between the two resonances and the signal--signal interference contribution and between the two signal--background interference contributions. Consequently, the total BSM contribution is significantly smaller than the individual pure resonance contributions. This scenario has mainly been chosen for illustrative purposes to demonstrate the impact of a proper treatment of large resonant-type mixing contributions in scenarios involving nearly mass-degenerate scalars (see also the discussion of \cref{fig:mass-deg-tot-compare}).

\cref{fig:nightmare-scenario} shows a scenario which also involves large cancellations between the individual contributions and a large destructive signal--signal interference contribution, giving rise to a total result for the $m(\ttb)$ distribution that lies within the experimental uncertainty band. The displayed scenario involves two $\CP$-mixed scalars with tree-level masses of $750$ and $766$ GeV. The total curve resulting from the sum of the resonance and interference contributions (the histograms are appropriately scaled with the respective K~factors) is seen to lie entirely within the statistical uncertainty band (indicated by the grey band). This scenario will therefore be difficult to discriminate from the background in the experimental analyses. We note that in order to obtain a reliable prediction for the total signal curve in the scenario displayed in \cref{fig:nightmare-scenario} an accurate theoretical prediction for the loop-level mixing effects of nearly mass-degenerate BSM scalars was required.

Next, we turn to a scenario for which the BSM contribution clearly exceeds the background uncertainty band. The scenario in \cref{fig:low-mass-resonance-scenario} has two $\CP$-mixed scalars with tree-level masses of $550$ and $600$ GeV. For this scenario, the signal--signal interference is smaller than the sum of the signal peaks. As a consequence of folding with the gluon PDFs, the limited experimental resolution (the corresponding plot with $0\%$ smearing is shown in \cref{fig:low-mass-0-smear} in \cref{appendix:shape-0-smear} for comparison) and the signal--background interferences, the $m(\ttb)$ distribution features a big broad peak at $m(\ttb) \sim 400\,\text{GeV}$ even though the masses of the BSM scalars are significantly larger (the distribution also has a small dip at $m(\ttb) \sim 670\,\text{GeV}$). This shows that the interpretation of a possible excess in the di-top searches would not necessarily point to a signal arising from the production of a single new particle with a mass near the observed peak. This motivates to look for complementary search channels like four-top production in order to determine the actual nature of a possibly observed signal.

While the scenarios discussed so far are not directly connected to concrete BSM models, we discuss in \cref{fig:full-c2hdm-bp3-scenario} a c2HDM scenario based on ``benchmark point 3'' of~\ccite{basler2020di}, which we mapped to our simplified model framework as outlined in \cref{c2hdm-intro}. The scenario involves two $\CP$-mixed scalars with tree-level masses of $608.6$ and $609.4$ GeV. For this scenario we find a similar behaviour as for the one of \cref{fig:low-mass-resonance-scenario}, demonstrating that as expected the patterns discussed for the scenarios above can indeed be realised in concrete BSM models. The signal--signal interference contribution is seen to be quite small in this scenario. While the individual signal contributions as expected have a peak near $600$~GeV, the large signal--background interference contributions in combination with the folding with the gluon PDFs give rise to a total result (blue curve in \cref{fig:full-c2hdm-bp3-scenario}) that has a peak just above the $t \bar t$ threshold. In fact, the comparison with the indicated experimental uncertainty band (grey band in \cref{fig:full-c2hdm-bp3-scenario}) shows that this enhancement of the cross-section near the $t \bar t$ threshold would be the only experimental signature of this scenario that could be resolved with the expected experimental sensitivity based on the luminosity of the full Run~3 of the LHC, while the dip in the $m(\ttb)$ distribution near $700$~GeV would be very difficult to distinguish from the background. Thus, the searches for BSM Higgs bosons in the $t \bar t$ final state at the LHC have the generic feature that the experimental sensitivity for discriminating BSM effects from the background tends to be highest in the $t \bar t$ threshold region. This emphasises the important conclusion that an observed excess near the $t \bar t$ threshold may not necessarily be caused by a new state with a mass close to the $t \bar t$ threshold or by a bound-state type effect at the threshold, but may actually occur as a consequence of the presence of new states with masses far above the $t \bar t$ threshold like in the scenario displayed in \cref{fig:full-c2hdm-bp3-scenario}.

We note that these findings are of interest in the context of the very significant excess of much more than $5\,\sigma$ above the perturbative QCD background that was recently observed by the CMS collaboration near the $t \bar t$ threshold in their analysis based on the full Run~2 data~\cite{CMS:2024ynj}. While we do not pursue here a dedicated analysis of the specific pattern that was observed by CMS, we emphasise that the mere fact that the observed excess occurs near the $t \bar t$ threshold by itself does not exclude the possibility that it could arise, possibly in combination with a bound-state type effect at the $t \bar t$ threshold, from BSM states with much higher masses. In fact, the current experimental limits leave more room for such BSM states at higher masses than it is the case for a single $\CP$-odd Higgs boson with a mass near the $t \bar t$ threshold.

	\section{Conclusions}
\label{sec:conclusions}
In this work, we investigated the impact of interference effects on BSM searches in the di-top final state. For our analysis, we employed a simplified model framework which extends the SM by two additional ($\CP$-admixed) scalars with masses above the di-top threshold and large couplings to top quarks, as they occur in many extensions of the SM Higgs sector.  The production of scalars via gluon--gluon fusion proceeding through a virtual top-quark loop and their subsequent decay to a pair of top quarks is affected by large interference effects with the SM di-top QCD background. Moreover, in the presence of more than one BSM scalar, signal--signal interference contributions can have a significant impact. These interference effects can significantly alter the invariant mass distribution and thereby affect the sensitivity of experimental searches targeting the di-top final state. 

For the first time, we studied the effect of loop-level mixing between the scalars in the di-top final state. This mixing between the tree-level mass eigenstates via loop-corrected propagators not only shifts the masses and decay widths of the involved particles but also affects their couplings to top quarks. We incorporated loop-level mixing using the $\bm{Z}$-matrix formalism. We found that the complex $\bm{Z}$-matrix elements introduce additional phase shifts between the interference contributions and thereby significantly modify the tree-level invariant mass distributions. Furthermore we demonstrated that the inclusion of Z~factors is crucial for regulating unphysical enhancements of the signal peaks that would occur if only the loop-corrected poles and decay widths were used. This clearly shows the importance of properly taking into account loop-level mixing effects in scenarios of extended Higgs sectors when confronting their predictions with experimental results.

Moreover, the Monte-Carlo implementation that we have provided should facilitate the incorporation of signal--signal interference and loop-level mixing contributions in the interpretation of experimental results. Emulating the effects of the limited experimental resolution by using a Gaussian smearing of the invariant mass of the top-quark pairs at the histogram level, we studied various scenarios with loop-level mixing at the hadronic level and compared the resulting patterns with the predictions at the parton level. We demonstrated that the limited experimental resolution washes out many of the features that would be present at the parton level and additional patterns arising from the folding with the gluon PDFs. As a consequence, we find that for the case of an observed excess in this search the limited resolution of the experimental analyses may make it difficult to determine the underlying origin of the detected excess. This is visible for the investigated scenarios with two $\CP$-mixed scalars since the resonance contributions of the two BSM scalars, their interferences with the QCD background and the signal--signal interference contribution combine in a non-trivial way to the overall distribution. In view of the limited experimental resolution it may be challenging to distinguish the case of a single new state from the case of two $\CP$-mixed states and to infer reliable information on the mass(es) of the produced BSM state(s).

As specific examples we investigated scenarios with large cancellations between the signal peaks and the signal--signal interference and demonstrated that the resulting overall distribution may be difficult to extract from the background. In addition, we demonstrated the versatility of the adopted model-independent approach by mapping the parameters in the neutral sector of the c2HDM to the Yukawa-coupling modifiers of our simplified model framework. Considering a scenario comprising two BSM Higgs bosons with masses far above the $t \bar t$ threshold, in this example at about $600$~GeV, we find that the combined effect of the interference contributions and the folding with the gluon PDFs gives rise to a signal that peaks at the $t \bar t$ threshold. The dip in the $m(\ttb)$ distribution occurring at a much higher value of $m(\ttb)$, in this case at about $700$~GeV, may be difficult to disentangle from the experimental uncertainty band. Thus, as a generic feature of searches in the $t \bar t$ final state at the LHC we find that even BSM states with much higher masses would manifest themselves in this search channel predominantly via the experimental signature of an excess of events in the $t \bar t$ threshold region (where the search has its highest sensitivity if the background is sufficiently well understood). We have discussed the implications of these results in the context of the recent excess above the perturbative QCD background that was observed by the CMS collaboration near the $t \bar t$ threshold. In particular, our results demonstrate that restricting possible interpretations of an excess at the $t \bar t$ threshold only to the options of a bound-state type effect at the threshold or a BSM state with a mass close to the $t \bar t$ threshold may be too restrictive, since BSM states with significantly higher masses could also contribute to an excess at the $t \bar t$ threshold.

Our study highlights the importance of interference contributions not only of signal--background but possibly also of signal--signal type for the interpretation of experimental searches in the di-top final states in extended Higgs sectors. For scenarios where such interference effects are large we expect that other search channels with different interference patterns --- e.g., four-top production --- will provide important complementary information. We leave detailed studies of the possible interplay between different search channels for future work.

	\section*{Acknowledgements}
\sloppy{
We thank Laurids Jeppe for helpful communications regarding various experimental aspects. Moreover, we thank Thomas Biek{\"o}tter, Prisco Lo Chiatto, Elina Fuchs, Philipp Gadow, Krisztian Peters, Panagiotis Stylianou, and Felix Yu for valuable discussions. H.B.\ acknowledges support by the Deutsche Forschungsgemeinschaft (DFG, German Research Foundation) under grant 396021762 – TRR 257: Particle
Physics Phenomenology after the Higgs Discovery. R.K.\ and G.W.\ acknowledge support by the Deutsche Forschungsgemeinschaft (DFG, German Research Foundation) under Germany’s Excellence Strategy – EXC 2121 “Quantum Universe” – 390833306. This work has been partially funded by the Deutsche Forschungsgemeinschaft (DFG, German Research Foundation) - 491245950.
}

    \begin{appendices}

\counterwithin{equation}{section}
\appendix

\section{Differential cross-section(s) with loop-level mixing}
\label{appendix:cross-section-z-fac}
Following the prescription as described in \cref{expr-ct-ctt-replacements-ct1,expr-ct-ctt-replacements-ctt1,expr-ct-ctt-replacements-ct2,expr-ct-ctt-replacements-ctt2} and taking into account the loop-corrected masses arising from the complex poles of the propagators, we obtain the following expressions
\allowdisplaybreaks
\begin{subequations}
\label{expr-dcs-two-mixed-replacement}
\begin{align}
	& \dfrac{\Dx\hat{\sigma}_\text{S}}{\Dx z} = \dfrac{3\alpha^2_\text{s}G^2_\text{F}\mt^2}{8192\pi^3}\hat{s}^2 \left(\dfrac{1}{(\hat{s} - M^2_{h_1})^2 + \Gamma^2_{h_1} M^2_{h_1}} \scaleobj{1.5}\times \right. \nonumber \\[2pt]
    & \hspace{50pt} \left(\left\vert\bm{Z}^{h_1}_{h_{1,\tree}}c_{t,1} + \bm{Z}^{h_1}_{h_{2,\tree}}c_{t,2}\right\vert^2\Bhat^3 + \left\vert\bm{Z}^{h_1}_{h_{1,\tree}}\tilde{c}_{t,1} + \bm{Z}^{h_1}_{h_{2,\tree}}\tilde{c}_{t,2}\right\vert^2\Bhat\right) \scaleobj{1.5}\times \nonumber \\[2pt]
    & \hspace{50pt} \left(\left\vert\left(\bm{Z}^{h_1}_{h_{1,\tree}}c_{t,1} + \bm{Z}^{h_1}_{h_{2,\tree}}c_{t,2}\right)A^{H}_{1/2}(\tau_t)\right\vert^2 + \left\vert\left(\bm{Z}^{h_1}_{h_{1,\tree}}\tilde{c}_{t,1} + \bm{Z}^{h_1}_{h_{2,\tree}}\tilde{c}_{t,2}\right)A^{A}_{1/2}(\tau_t)\right\vert^2\right) \nonumber \\[3pt]
	\label{expr-dcs-two-mixed-sig-zfac} & \hspace{50pt} \scaleobj{1.4}{+} \hspace{10pt} \dfrac{1}{(\hat{s} - M^2_{h_2})^2 + \Gamma^2_{h_2} M^2_{h_2}} \scaleobj{1.5}\times \nonumber \\[2pt]
    & \hspace{50pt} \left(\left\vert\bm{Z}^{h_2}_{h_{2,\tree}}c_{t,2} + \bm{Z}^{h_2}_{h_{1,\tree}}c_{t,1}\right\vert^2\Bhat^3 + \left\vert\bm{Z}^{h_2}_{h_{2,\tree}}\tilde{c}_{t,2} + \bm{Z}^{h_2}_{h_{1,\tree}}\tilde{c}_{t,1}\right\vert^2\Bhat\right) \scaleobj{1.5}\times \nonumber \\[2pt]
    & \hspace{50pt} \left(\left\vert\left(\bm{Z}^{h_2}_{h_{2,\tree}}c_{t,2} + \bm{Z}^{h_2}_{h_{1,\tree}}c_{t,1}\right)A^{H}_{1/2}(\tau_t)\right\vert^2 + \left\vert\left(\bm{Z}^{h_2}_{h_{2,\tree}}\tilde{c}_{t,2} + \bm{Z}^{h_2}_{h_{1,\tree}}\tilde{c}_{t,1}\right)A^{A}_{1/2}(\tau_t)\right\vert^2\right) \nonumber \\[2pt]
	& \hspace{50pt} \scaleobj{1.4}{+} \ 2\times\text{Re}\Biggl[\dfrac{1}{(\hat{s} - M^2_{h_1} + i\Gamma_{h_1} M_{h_1}) \cdot (\hat{s} - M^2_{h_2} - i\Gamma_{h_2} M_{h_2})} \scaleobj{1.5}\times \\
    & \hspace{50pt} \left(\left(\bm{Z}^{h_1}_{h_{1,\tree}}c_{t,1} + \bm{Z}^{h_1}_{h_{2,\tree}}c_{t,2}\right)\left(\bm{Z}_{h_{2,\tree}}^{h_2,\scaleobj{1.5}*}c_{t,2} + \bm{Z}_{h_{1,\tree}}^{h_2,\scaleobj{1.5}*}c_{t,1}\right)\Bhat^3 \right. \nonumber \\[2pt]
    & \hspace{45pt} 
    \left.\left. \scaleobj{1.4}{+} \left(\bm{Z}^{h_1}_{h_{1,\tree}}\tilde{c}_{t,1} + \bm{Z}^{h_1}_{h_{2,\tree}}\tilde{c}_{t,2}\right)\left(\bm{Z}_{h_{2,\tree}}^{h_2,\scaleobj{1.5}*}\tilde{c}_{t,2} + \bm{Z}_{h_{1,\tree}}^{h_2,\scaleobj{1.5}*}\tilde{c}_{t,1}\right)\Bhat\right) \cdot \Delta(1,2) \Biggr]\right) \,, \nonumber \\[3pt]
	& \dfrac{\Dx\hat{\sigma}_\text{I}}{\Dx z} = -\dfrac{\alpha^2_\text{s}G_\text{F}\mt^2}{64\sqrt{2}\,\pi}\dfrac{1}{1-\Bhat^2 z^2}\ \scaleobj{1.5}\times \nonumber \\[4pt]
	\label{expr-dcs-two-mixed-intf-zfac} & \hspace{30pt} \text{Re}\left[\dfrac{\left(\bm{Z}^{h_1}_{h_{1,\tree}}c_{t,1} + \bm{Z}^{h_1}_{h_{2,\tree}}c_{t,2}\right)^2\Bhat^3 A^{H}_{1/2}(\tau_t) + \left(\bm{Z}^{h_1}_{h_{1,\tree}}\tilde{c}_{t,1} + \bm{Z}^{h_1}_{h_{2,\tree}}\tilde{c}_{t,2}\right)^2\Bhat A^{A}_{1/2}(\tau_t)}{\hat{s} - M^2_{h_1} + i\Gamma_{h_1} M_{h_1}}\right. \nonumber \\
    & \hspace{30pt} \left. \scaleobj{1.4}{+} \dfrac{\left(\bm{Z}^{h_2}_{h_{2,\tree}}c_{t,2} + \bm{Z}^{h_2}_{h_{1,\tree}}c_{t,1}\right)^2\Bhat^3 A^{H}_{1/2}(\tau_t) + \left(\bm{Z}^{h_2}_{h_{2,\tree}}\tilde{c}_{t,2} + \bm{Z}^{h_2}_{h_{1,\tree}}\tilde{c}_{t,1}\right)^2\Bhat A^{A}_{1/2}(\tau_t)}{\hat{s} - M^2_{h_2} + i\Gamma_{h_2} M_{h_2}} \right] \,,
\intertext{where}
	\label{expr-two-scalars-delta-zfac} & \Delta(1,2) = \left(\left(\bm{Z}^{h_1}_{h_{1,\tree}}c_{t,1} + \bm{Z}^{h_1}_{h_{2,\tree}}c_{t,2}\right)A^{H}_{1/2}(\tau_t)\left(\bm{Z}_{h_{2,\tree}}^{h_2,\scaleobj{1.5}*}c_{t,2} + \bm{Z}_{h_{1,\tree}}^{h_2,\scaleobj{1.5}*}c_{t,1}\right)A^{H,\scaleobj{1.5}*}_{1/2}(\tau_t) \right. \nonumber \\ 
    & \hspace{50pt} \scaleobj{1.4}{+} \left. \left(\bm{Z}^{h_1}_{h_{1,\tree}}\tilde{c}_{t,1} + \bm{Z}^{h_1}_{h_{2,\tree}}\tilde{c}_{t,2}\right)A^{A}_{1/2}(\tau_t) \left(\bm{Z}_{h_{2,\tree}}^{h_2,\scaleobj{1.5}*}\tilde{c}_{t,2} + \bm{Z}_{h_{1,\tree}}^{h_2,\scaleobj{1.5}*}\tilde{c}_{t,1}\right)A^{A,\scaleobj{1.5}*}_{1/2}(\tau_t)\right) \,, \nonumber \\
	& \text{and } \tau_t = \dfrac{\hat{s}}{4\mt^2} \,. 
\end{align}
\end{subequations}
As before, ``S'' denotes the signal and ``I'' denotes the signal--background interference contribution (Sig-Bkg Intf.). In the signal contribution, the term ($2\times\text{Re}[\cdots]$) corresponds to the ``signal--signal'' (or, ``sig$1$-sig$2$'') interference term.

\section{\texorpdfstring{$\overline{\text{MS}}$}{MS}-renormalised self-energies}
\label{appendix:ms-self-energy}
The $\overline{\text{MS}}$-renormalised self-energies are given by
\begin{flalign}
    \hat{\Sigma}_{h_i h_i}(p^2) = \dfrac{3\alpha \mt^2}{8\pi M_W^2 \sin^2\theta_W} & \left[-2\left(\cti^2 + \ctti^2\right)A_0^{\text{fin.}}(\mt^2) \right. && \nonumber \\
    & \scaleobj{1.1}{+} \left. \left(\cti^2\left(p^2 - 4\mt^2\right) + \ctti^2 p^2\right)B_0^{\text{fin.}}(p^2, \mt^2, \mt^2)\right] \,, &&
\end{flalign}
and
\begin{flalign}
     \hat{\Sigma}_{h_i h_j}(p^2) = \dfrac{3\alpha \mt^2}{8\pi M_W^2 \sin^2\theta_W} & \left[-2\left(\cti\ctj + \ctti\cttj\right)A_0^{\text{fin.}}(\mt^2) \right. && \nonumber \\
     & \scaleobj{1.1}{+} \left. \left(\cti\ctj\left(p^2 - 4\mt^2\right) + \ctti\cttj p^2\right)B_0^{\text{fin.}}(p^2, \mt^2, \mt^2)\right] \,,
\end{flalign}
where $A_0$ and $B_0$ are the standard scalar one-loop integrals as defined in~\ccite{Denner:1991kt}, and ``fin.'' denotes their UV-finite parts.

\section{UFO model file and generation of Monte-Carlo events}
\label{appendix:ufo-details}
Our \textit{Universal FeynRules Output} (UFO) model file incorporates two $\CP$-mixed scalars that we label as $S_1$ and $S_2$ in the model file. The full top-triangle loop is implemented in the model file using a Fortran routine which defines the corresponding loop functions for producing a $\CP$-even or a $\CP$-odd scalar. We validated our UFO model file by comparing it with the publicly available UFO model file~\cite{mg-ggH-ggA-loop}, on which our implementation is based, for the case without $\CP$~violation.

To facilitate event generation the model file implements the two coupling orders \texttt{QS1}, which denotes the coupling of $S_1$, and \texttt{QS2}, which denotes the coupling of $S_2$. This allows simulating the individual contributions to the di-top processes separately as follows:
\begin{center}
\begin{lstlisting}
$\commentref{QCD background}$: g g > t t~ QS1^2==0 QS2^2==0
$\commentref{signal~1 (resonance)}$: g g > t t~ QS1^2==4 QS2^2==0
$\commentref{signal~2 (resonance)}$: g g > t t~ QS1^2==0 QS2^2==4
$\commentref{signal~1 -- background (interference)}$: g g > t t~ QS1^2==2 QS2^2==0
$\commentref{signal~2 -- background (interference)}$: g g > t t~ QS1^2==0 QS2^2==2
$\commentref{signal~1 -- signal~2 (interference)}$: g g > t t~ QS1^2==2 QS2^2==2
\end{lstlisting}
\end{center}

\section{Input parameters used for the overview scenarios}
\label{appendix:sketch-parameters}
The tree-level masses, decay widths, Yukawa coupling modifiers, Z~factors, and the K~factors for the scenarios considered in \cref{fig:lineshape-sketch-15,fig:lineshape-sketch-0} are tabulated in \cref{table:parameters-1,table:parameters-2}.
\begin{table}[!htb]
\begin{center}
    \begin{tabular}{|c|c|c|}
        \hline
        Scenario & (a) & (b) \\ \hline
        $M_{h_{1,\tree}}\ (M_{h_1})$ [GeV] & $750\ (747.60)$ & $550\ (541.28)$ \\ \cline{1-1}
        $\Gamma_{h_{1}}$ [GeV] & $24.50$ & $27.48$ \\ \cline{1-1}
        $c_{t,1}$ & $-0.4$ & $0.8$ \\ \cline{1-1}
        $\tilde{c}_{t,1}$ & $0.5$ & $0.7$ \\ \hline
        $M_{h_{2,\tree}}\ (M_{h_2})$ [GeV] & $766\ (751.48)$ & $600\ (577.69)$ \\ \cline{1-1}
        $\Gamma_{h_{2}}$ [GeV] & $19.36$ & $38.78$ \\ \cline{1-1}
        $c_{t,2}$ & $0.4$ & $0.4$ \\ \cline{1-1}
        $\tilde{c}_{t,2}$ & $0.8$ & $-1.3$ \\ \hline
        $\bm{Z}$-matrix 
        & $\begin{pmatrix} 1.12 - 0.59i & 0.82 + 0.75i \\ -0.83 - 0.78i & 1.1 - 0.58i \end{pmatrix}$ 
        & $\begin{pmatrix} 1.00 - 0.08i & -0.19 - 0.23i \\ 0.20 + 0.27i & 0.99 - 0.09i \end{pmatrix}$ \\ \hline
        $K(\text{sig}(1))$ & $2.49$ & $2.60$ \\ \cline{1-1}
        $K(\text{sig}(2))$ & $2.48$ & $2.54$ \\ \hline
    \end{tabular}
    \caption{Parameters that are used for scenarios (a) and (b) in the overview plots in \cref{fig:lineshape-sketch-15,fig:lineshape-sketch-0}.}
    \label{table:parameters-1}
\end{center}
\end{table}
\begin{table}[!htb]
\begin{center}
    \begin{tabular}{|c|c|c|}
        \hline
        Scenario & (c) & (d) \\ \hline
        $M_{h_{1,\tree}}\ (M_{h_1})$ [GeV] & $500\ (494.87)$ & $550\ (544.83)$ \\ \cline{1-1}
        $\Gamma_{h_{1}}$ [GeV] & $12.13$ & $32.42$ \\ \cline{1-1}
        $c_{t,1}$ & $0.0$ & $0.8$ \\ \cline{1-1}
        $\tilde{c}_{t,1}$ & $0.7$ & $0.5$ \\ \hline
        $M_{h_{2,\tree}}\ (M_{h_2})$ [GeV] & $600\ (581.20)$ & $580\ (560.67)$ \\ \cline{1-1}
        $\Gamma_{h_{2}}$ [GeV] & $39.66$ & $19.12$ \\ \cline{1-1}
        $c_{t,2}$ & $1.0$ & $0.4$ \\ \cline{1-1}
        $\tilde{c}_{t,2}$ & $1.0$ & $1.2$ \\ \hline
        $\bm{Z}$-matrix 
        & $\begin{pmatrix} 1.00 - 0.02i & 0.07 + 0.07i \\ -0.09 - 0.12i & 0.97 - 0.06i \end{pmatrix}$ 
        & $\begin{pmatrix} 0.95 - 0.31i & 0.57 + 0.40i \\ -0.61 - 0.45i & 0.93 - 0.29i \end{pmatrix}$ \\ \hline
        $K(\text{sig}(1))$ & $2.63$ & $2.57$ \\ \cline{1-1}
        $K(\text{sig}(2))$ & $2.50$ & $2.55$ \\ \hline
    \end{tabular}
    \caption{Parameters that are used for scenarios (c) and (d) in the overview plots in \cref{fig:lineshape-sketch-15,fig:lineshape-sketch-0}.}
    \label{table:parameters-2}
\end{center}
\end{table}

\section{Di-top invariant mass distribution with no Gaussian smearing}
\label{appendix:shape-0-smear}
\begin{figure}[!htb]
    \centering
    \includegraphics[width=0.95\textwidth,keepaspectratio]{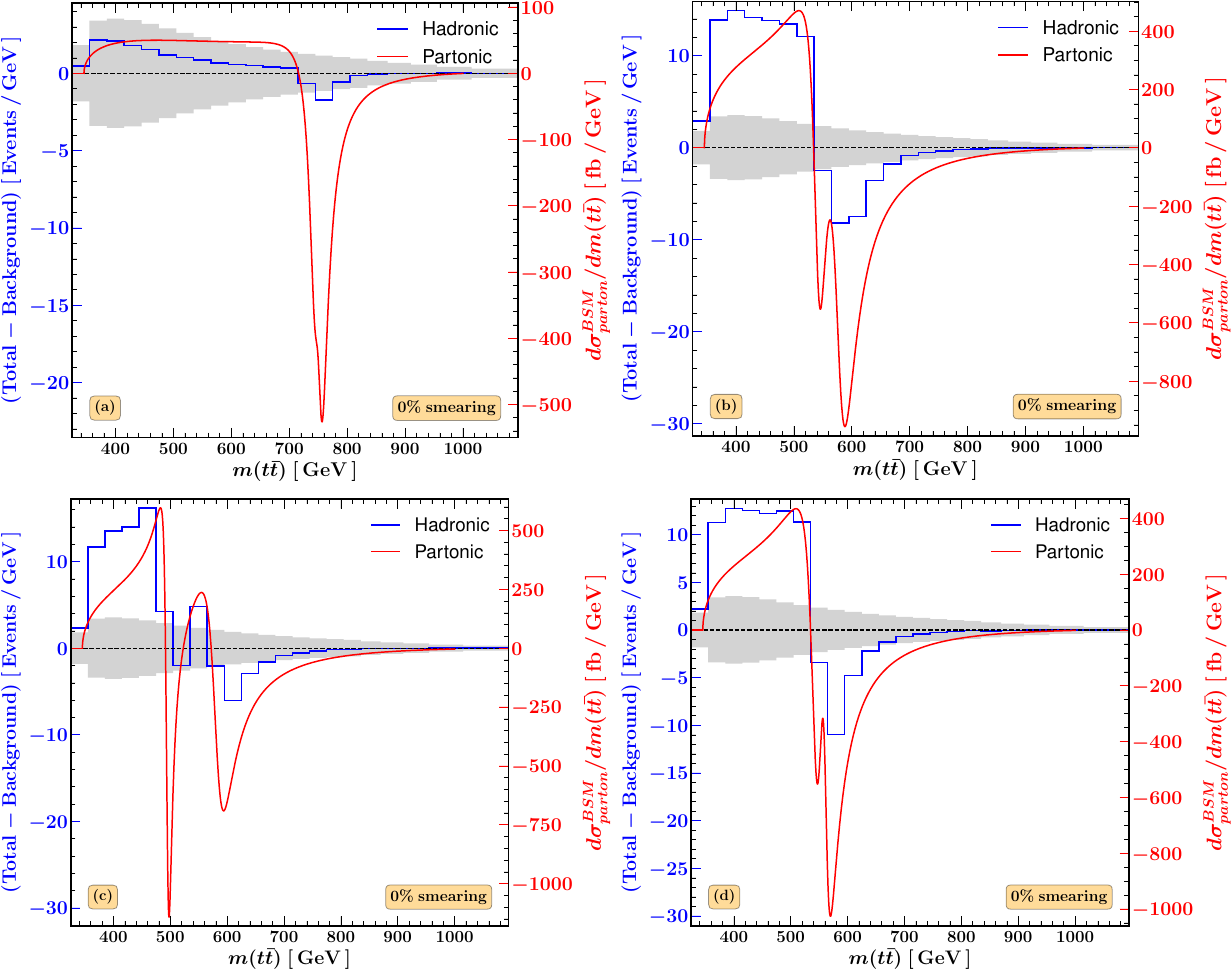}
    \caption{(a)--(d) Background-subtracted $m(\ttb)$ distributions resulting from the sum of all resonance and interference contributions. The plots overlay the partonic-level (shown in red) and the hadronic-level (shown in blue) expectations in the $m(\ttb)$ distribution. The hadronic-level plots are shown for a Gaussian smearing of $0\%$, i.e., for the case where detector effects are neglected yielding an exact reconstruction of the invariant mass, $m(\ttb)$. The grey band denotes the statistical uncertainty on the SM $\ttb$ background at the hadronic level. The input parameters used for obtaining the plots in (a)--(d) are listed in \cref{appendix:sketch-parameters}.}
    \label{fig:lineshape-sketch-0}
\end{figure}
We present in \cref{fig:lineshape-sketch-0} the background-subtracted $m(\ttb)$ distributions both at the partonic (red curves) and hadronic level (blue curves) for the four illustrative scenarios of \cref{fig:lineshape-sketch-15} with $0\%$ smearing. The grey band is the statistical uncertainty of the SM QCD background at the hadronic level. The input parameters used for obtaining the various partonic-level and hadronic-level curves in \cref{fig:lineshape-sketch-0} arise are given in \cref{appendix:sketch-parameters}.

\begin{figure}[!htb]
    \centering
    \includegraphics[height=0.95\linewidth,width=0.95\linewidth,keepaspectratio]{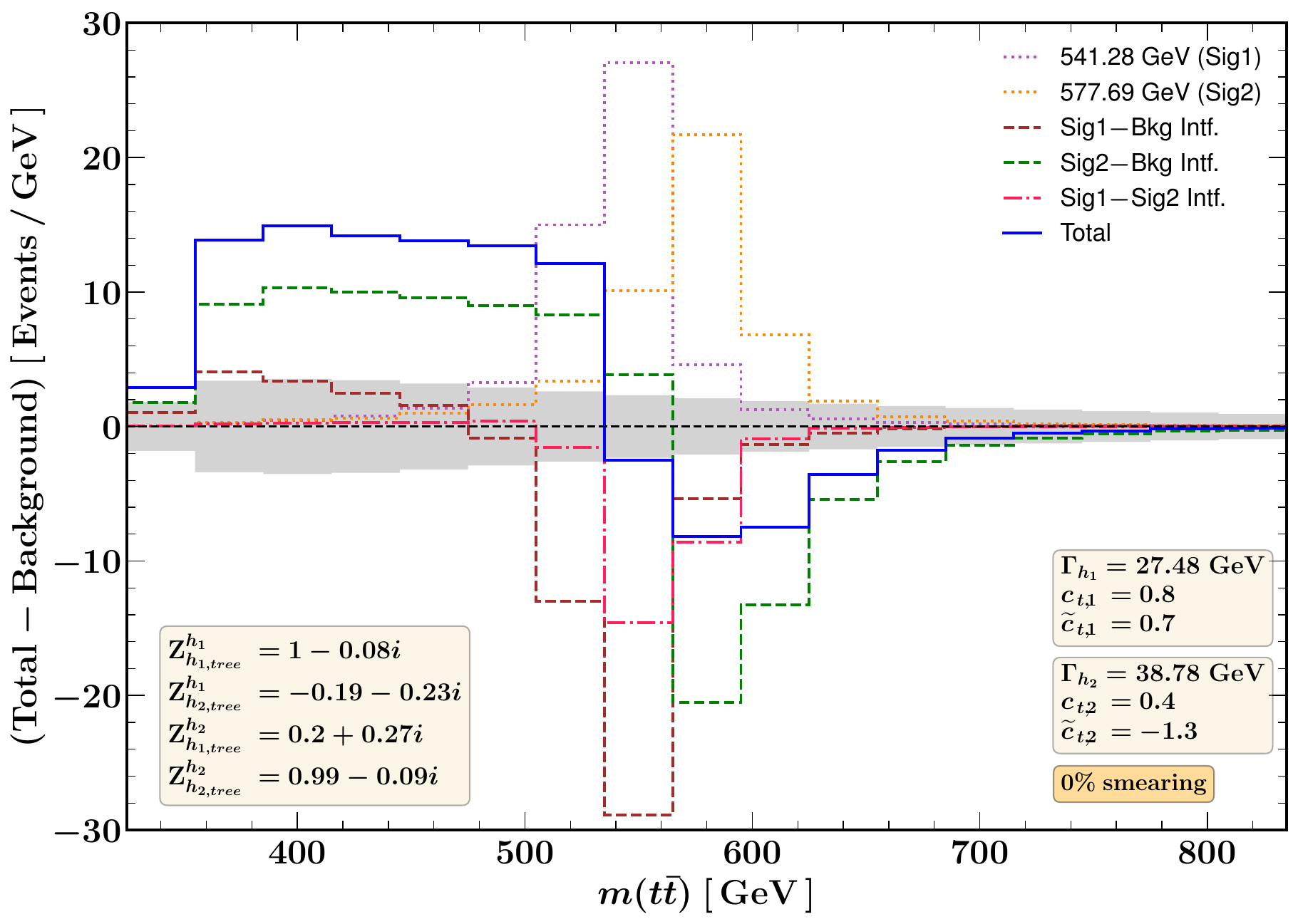}
    \caption{Close-in-mass scenario: contributions to the background-subtracted $m(\ttb)$ distribution at the hadronic level for the same scenario as in \cref{fig:low-mass-resonance-scenario} (the tree-level masses are $550$ GeV and $600$ GeV), but with $0\%$ smearing.}
    \label{fig:low-mass-0-smear}
\end{figure}
Moreover, we show in \cref{fig:low-mass-0-smear} the scenario of \cref{fig:low-mass-resonance-scenario} without smearing. The impact of the folding with the gluon PDFs on the peak--dip structure arising from the signal--background interference is clearly visible in this case. It gives rise to a plateau-like shape of the peak between the $t \bar t$ threshold and the mass of the produced scalar(s) (in this case about $550$~GeV). The effect of the $15\%$ Gaussian smearing that is applied in \cref{fig:low-mass-resonance-scenario} in order to account for the limited detector resolution is seen to reduce the significance of the peak and the dip in comparison to the experimental uncertainty band. While the dip at about 670~GeV would be difficult to experimentally resolve in this case, the plateau-like shape of the peak is modified into a broad peak at about $400$~GeV in \cref{fig:low-mass-resonance-scenario} as a result of the applied smearing.

We conclude from this investigation that an ideal detector with $0\%$ smearing in the $m(\ttb)$ resolution would preserve several of the distinct features of the parton-level distributions and clearly display the effect of an enhancement in the $t \bar t$ threshold region arising from the folding with the gluon PDFs. The limited experimental resolution of the actual analyses may make it difficult to disentangle the underlying origin of an observed excess, which is exemplified for the considered case of two $\CP$-mixed scalars where the resonance contributions of the two BSM scalars, their interferences with the QCD background and the signal--signal interference contribution combine in a non-trivial way to the overall distribution.

\end{appendices}
    \clearpage
    \addcontentsline{toc}{section}{References}        
    \bibliography{references}

\providecommand{\href}[2]{#2}\begingroup\raggedright\begin{thebibliography}{10}

\bibitem{CMS:2012qbp}
{\scshape CMS} collaboration, \emph{{Observation of a New Boson at a Mass of
  125 GeV with the CMS Experiment at the LHC}},
  \href{https://doi.org/10.1016/j.physletb.2012.08.021}{\emph{Phys. Lett. B}
  {\bfseries 716} (2012) 30} [\href{https://arxiv.org/abs/1207.7235}{{\ttfamily
  1207.7235}}].

\bibitem{ATLAS:2012yve}
{\scshape ATLAS} collaboration, \emph{{Observation of a new particle in the
  search for the Standard Model Higgs boson with the ATLAS detector at the
  LHC}}, \href{https://doi.org/10.1016/j.physletb.2012.08.020}{\emph{Phys.
  Lett. B} {\bfseries 716} (2012) 1}
  [\href{https://arxiv.org/abs/1207.7214}{{\ttfamily 1207.7214}}].

\bibitem{Branco:2011iw}
G.C.~Branco, P.M.~Ferreira, L.~Lavoura, M.N.~Rebelo, M.~Sher and J.P.~Silva,
  \emph{{Theory and phenomenology of two-Higgs doublet models}},
  \href{https://doi.org/10.1016/j.physrep.2012.02.002}{\emph{Phys. Rept.}
  {\bfseries 516} (2012) 1} [\href{https://arxiv.org/abs/1106.0034}{{\ttfamily
  1106.0034}}].

\bibitem{Ivanov:2017dad}
I.P.~Ivanov, \emph{{Building and testing models with extended Higgs sectors}},
  \href{https://doi.org/10.1016/j.ppnp.2017.03.001}{\emph{Prog. Part. Nucl.
  Phys.} {\bfseries 95} (2017) 160}
  [\href{https://arxiv.org/abs/1702.03776}{{\ttfamily 1702.03776}}].

\bibitem{gunion2018higgs}
J.F.~Gunion, H.E.~Haber, G.L.~Kane and S.~Dawson, \emph{{The Higgs Hunter's
  Guide}}, Vol.~80, CRC Press (2018),
  \texttt{doi:}\href{https://doi.org/10.1201/9780429496448}{10.1201/9780429496448}.

\bibitem{Gaemers:1984sj}
K.J.F.~Gaemers and F.~Hoogeveen, \emph{{Higgs Production and Decay Into Heavy
  Flavors With the Gluon Fusion Mechanism}},
  \href{https://doi.org/10.1016/0370-2693(84)91711-8}{\emph{Phys. Lett. B}
  {\bfseries 146} (1984) 347}.

\bibitem{Dicus:1994bm}
D.~Dicus, A.~Stange and S.~Willenbrock, \emph{{Higgs decay to top quarks at
  hadron colliders}},
  \href{https://doi.org/10.1016/0370-2693(94)91017-0}{\emph{Phys. Lett. B}
  {\bfseries 333} (1994) 126}
  [\href{https://arxiv.org/abs/hep-ph/9404359}{{\ttfamily hep-ph/9404359}}].

\bibitem{Frederix:2007gi}
R.~Frederix and F.~Maltoni, \emph{{Top pair invariant mass distribution: A
  Window on new physics}},
  \href{https://doi.org/10.1088/1126-6708/2009/01/047}{\emph{JHEP} {\bfseries
  01} (2009) 047} [\href{https://arxiv.org/abs/0712.2355}{{\ttfamily
  0712.2355}}].

\bibitem{Craig:2015jba}
N.~Craig, F.~D'Eramo, P.~Draper, S.~Thomas and H.~Zhang, \emph{{The Hunt for
  the Rest of the Higgs Bosons}},
  \href{https://doi.org/10.1007/JHEP06(2015)137}{\emph{JHEP} {\bfseries 06}
  (2015) 137} [\href{https://arxiv.org/abs/1504.04630}{{\ttfamily
  1504.04630}}].

\bibitem{Jung:2015gta}
S.~Jung, J.~Song and Y.W.~Yoon, \emph{{Dip or nothingness of a Higgs resonance
  from the interference with a complex phase}},
  \href{https://doi.org/10.1103/PhysRevD.92.055009}{\emph{Phys. Rev. D}
  {\bfseries 92} (2015) 055009}
  [\href{https://arxiv.org/abs/1505.00291}{{\ttfamily 1505.00291}}].

\bibitem{Bernreuther:2015fts}
W.~Bernreuther, P.~Galler, C.~Mellein, Z.G.~Si and P.~Uwer, \emph{{Production
  of heavy Higgs bosons and decay into top quarks at the LHC}},
  \href{https://doi.org/10.1103/PhysRevD.93.034032}{\emph{Phys. Rev. D}
  {\bfseries 93} (2016) 034032}
  [\href{https://arxiv.org/abs/1511.05584}{{\ttfamily 1511.05584}}].

\bibitem{Carena:2016npr}
M.~Carena and Z.~Liu, \emph{{Challenges and opportunities for heavy scalar
  searches in the $ t\overline{t} $ channel at the LHC}},
  \href{https://doi.org/10.1007/JHEP11(2016)159}{\emph{JHEP} {\bfseries 11}
  (2016) 159} [\href{https://arxiv.org/abs/1608.07282}{{\ttfamily
  1608.07282}}].

\bibitem{Hespel:2016qaf}
B.~Hespel, F.~Maltoni and E.~Vryonidou, \emph{{Signal background interference
  effects in heavy scalar production and decay to a top-anti-top pair}},
  \href{https://doi.org/10.1007/JHEP10(2016)016}{\emph{JHEP} {\bfseries 10}
  (2016) 016} [\href{https://arxiv.org/abs/1606.04149}{{\ttfamily
  1606.04149}}].

\bibitem{BuarqueFranzosi:2017jrj}
D.~Buarque~Franzosi, E.~Vryonidou and C.~Zhang, \emph{{Scalar production and
  decay to top quarks including interference effects at NLO in QCD in an EFT
  approach}}, \href{https://doi.org/10.1007/JHEP10(2017)096}{\emph{JHEP}
  {\bfseries 10} (2017) 096}
  [\href{https://arxiv.org/abs/1707.06760}{{\ttfamily 1707.06760}}].

\bibitem{Djouadi:2019cbm}
A.~Djouadi, J.~Ellis, A.~Popov and J.~Quevillon, \emph{{Interference effects in
  $ t\overline{t} $ production at the LHC as a window on new physics}},
  \href{https://doi.org/10.1007/JHEP03(2019)119}{\emph{JHEP} {\bfseries 03}
  (2019) 119} [\href{https://arxiv.org/abs/1901.03417}{{\ttfamily
  1901.03417}}].

\bibitem{RK:2022thesis}
R.~Kumar, \emph{{A model-independent analysis of interference effects in the $t
  \bar t$ final state at the LHC involving two CP-mixed Higgs bosons}},
  Master's thesis, University of Hamburg, 2022.

\bibitem{Bahl:2024fjb}
H.~Bahl, R.~Kumar and G.~Weiglein, \emph{{Analysis of interference effects in
  the di-top final state for CP-mixed scalars in extended Higgs sectors}},
  \href{https://doi.org/10.22323/1.449.0057}{\emph{PoS} {\bfseries EPS-HEP2023}
  (2024) 057}.

\bibitem{ATLAS:2017snw}
{\scshape ATLAS} collaboration, \emph{{Search for Heavy Higgs Bosons $A/H$
  Decaying to a Top Quark Pair in $pp$ Collisions at $\sqrt{s}=8\text{ }\text{
  }\mathrm{TeV}$ with the ATLAS Detector}},
  \href{https://doi.org/10.1103/PhysRevLett.119.191803}{\emph{Phys. Rev. Lett.}
  {\bfseries 119} (2017) 191803}
  [\href{https://arxiv.org/abs/1707.06025}{{\ttfamily 1707.06025}}].

\bibitem{CMS:2019pzc}
{\scshape CMS} collaboration, \emph{{Search for heavy Higgs bosons decaying to
  a top quark pair in proton-proton collisions at $\sqrt{s} =$ 13 TeV}},
  \href{https://doi.org/10.1007/JHEP04(2020)171}{\emph{JHEP} {\bfseries 04}
  (2020) 171} [\href{https://arxiv.org/abs/1908.01115}{{\ttfamily
  1908.01115}}].

\bibitem{Ginzburg:2002wt}
I.F.~Ginzburg, M.~Krawczyk and P.~Osland, \emph{{Two Higgs doublet models with
  CP violation}},  in \emph{{International Workshop on Linear Colliders (LCWS
  2002)}}, pp.~703--706, 11, 2002
  [\href{https://arxiv.org/abs/hep-ph/0211371}{{\ttfamily hep-ph/0211371}}].

\bibitem{Khater:2003wq}
W.~Khater and P.~Osland, \emph{{CP violation in top quark production at the LHC
  and two Higgs doublet models}},
  \href{https://doi.org/10.1016/S0550-3213(03)00300-6}{\emph{Nucl. Phys. B}
  {\bfseries 661} (2003) 209}
  [\href{https://arxiv.org/abs/hep-ph/0302004}{{\ttfamily hep-ph/0302004}}].

\bibitem{fontes2018c2hdm}
D.~Fontes, M.~M\"uhlleitner, J.C.~Rom{\~a}o, R.~Santos, J.P.~Silva and
  J.~Wittbrodt, \emph{{The C2HDM revisited}},
  \href{https://doi.org/10.1007/JHEP02(2018)073}{\emph{JHEP} {\bfseries 02}
  (2018) 073} [\href{https://arxiv.org/abs/1711.09419}{{\ttfamily
  1711.09419}}].

\bibitem{Fuchs:2014ola}
E.~Fuchs, S.~Thewes and G.~Weiglein, \emph{{Interference effects in BSM
  processes with a generalised narrow-width approximation}},
  \href{https://doi.org/10.1140/epjc/s10052-015-3472-z}{\emph{Eur. Phys. J. C}
  {\bfseries 75} (2015) 254} [\href{https://arxiv.org/abs/1411.4652}{{\ttfamily
  1411.4652}}].

\bibitem{Fuchs:2016swt}
E.~Fuchs and G.~Weiglein, \emph{{Breit-Wigner approximation for propagators of
  mixed unstable states}},
  \href{https://doi.org/10.1007/JHEP09(2017)079}{\emph{JHEP} {\bfseries 09}
  (2017) 079} [\href{https://arxiv.org/abs/1610.06193}{{\ttfamily
  1610.06193}}].

\bibitem{Fuchs:2017wkq}
E.~Fuchs and G.~Weiglein, \emph{{Impact of CP-violating interference effects on
  MSSM Higgs searches}},
  \href{https://doi.org/10.1140/epjc/s10052-018-5543-4}{\emph{Eur. Phys. J. C}
  {\bfseries 78} (2018) 87} [\href{https://arxiv.org/abs/1705.05757}{{\ttfamily
  1705.05757}}].

\bibitem{Bagnaschi:2018ofa}
E.~Bagnaschi et~al., \emph{{MSSM Higgs Boson Searches at the LHC: Benchmark
  Scenarios for Run~2 and Beyond}},
  \href{https://doi.org/10.1140/epjc/s10052-019-7114-8}{\emph{Eur. Phys. J. C}
  {\bfseries 79} (2019) 617}
  [\href{https://arxiv.org/abs/1808.07542}{{\ttfamily 1808.07542}}].

\bibitem{ATLAS:2024itc}
{\scshape ATLAS} collaboration, \emph{{ATLAS searches for additional scalars
  and exotic Higgs boson decays with the LHC Run~2 dataset}},
  [\href{https://arxiv.org/abs/2405.04914}{{\ttfamily 2405.04914}}].

\bibitem{ATLAS:2024vxm}
{\scshape ATLAS} collaboration, \emph{{Search for heavy neutral Higgs bosons
  decaying into a top quark pair in 140 fb$^{-1}$ of proton-proton collision
  data at $ \sqrt{s} $ = 13 TeV with the ATLAS detector}},
  \href{https://doi.org/10.1007/JHEP08(2024)013}{\emph{JHEP} {\bfseries 08}
  (2024) 013} [\href{https://arxiv.org/abs/2404.18986}{{\ttfamily
  2404.18986}}].

\bibitem{CMS:2024ynj}
{CMS collaboration}, \emph{{Search for heavy pseudoscalar and scalar bosons
  decaying to top quark pairs in proton-proton collisions at $\sqrt{s} = 13$
  TeV}},  \href{https://cds.cern.ch/record/2911775}{CMS-PAS-HIG-22-013}.
\newblock \url{https://cds.cern.ch/record/2911775}.

\bibitem{Biekotter:2021qbc}
T.~Biek\"otter, A.~Grohsjean, S.~Heinemeyer, C.~Schwanenberger and G.~Weiglein,
  \emph{{Possible indications for new Higgs bosons in the reach of the LHC:
  N2HDM and NMSSM interpretations}},
  \href{https://doi.org/10.1140/epjc/s10052-022-10099-1}{\emph{Eur. Phys. J. C}
  {\bfseries 82} (2022) 178}
  [\href{https://arxiv.org/abs/2109.01128}{{\ttfamily 2109.01128}}].

\bibitem{Anuar:2024qsz}
A.~Anuar, A.~Biek\"otter, T.~Biek\"otter, A.~Grohsjean, S.~Heinemeyer, L.~Jeppe
  et~al., \emph{{ALP-ine quests at the LHC: hunting axion-like particles via
  peaks and dips in $ t\overline{t} $ production}},
  \href{https://doi.org/10.1007/JHEP12(2024)197}{\emph{JHEP} {\bfseries 12}
  (2024) 197} [\href{https://arxiv.org/abs/2404.19014}{{\ttfamily
  2404.19014}}].

\bibitem{Arco:2025ydq}
F.~Arco, T.~Biek\"otter, P.~Stylianou and G.~Weiglein, \emph{{Top-quark spin
  correlations as a tool to distinguish pseudoscalar $A \to ZH$ and scalar $H
  \to ZA$ signatures in $Z t \bar t$ final states at the LHC}},
  [\href{https://arxiv.org/abs/2502.03443}{{\ttfamily 2502.03443}}].

\bibitem{Lu:2024twj}
C.-T.~Lu, K.~Cheung, D.~Kim, S.~Lee and J.~Song, \emph{{Can a pseudoscalar with
  a mass of 365 GeV in two-Higgs doublet models explain the CMS $t\bar{t}$
  excess?}}, \href{https://doi.org/10.1016/j.physletb.2024.139121}{\emph{Phys.
  Lett. B} {\bfseries 859} (2024) 139121}
  [\href{https://arxiv.org/abs/2410.08609}{{\ttfamily 2410.08609}}].

\bibitem{Fadin:1990wx}
V.S.~Fadin, V.A.~Khoze and T.~Sjostrand, \emph{{On the Threshold Behavior of
  Heavy Top Production}}, \href{https://doi.org/10.1007/BF01614696}{\emph{Z.
  Phys. C} {\bfseries 48} (1990) 613}.

\bibitem{Fadin:1991zw}
V.S.~Fadin and V.A.~Khoze, \emph{{Production of a pair of $t \bar{t}$ quarks
  near threshold}}, {\emph{Sov. J. Nucl. Phys.} {\bfseries 53} (1991) 692}.

\bibitem{Hoang:2000yr}
A.H.~Hoang et~al., \emph{{Top - anti-top pair production close to threshold:
  Synopsis of recent NNLO results}},
  \href{https://doi.org/10.1007/s1010500c0003}{\emph{Eur. Phys. J. direct}
  {\bfseries 2} (2000) 3}
  [\href{https://arxiv.org/abs/hep-ph/0001286}{{\ttfamily hep-ph/0001286}}].

\bibitem{Kiyo:2008bv}
Y.~Kiyo, J.H.~Kuhn, S.~Moch, M.~Steinhauser and P.~Uwer, \emph{{Top-quark pair
  production near threshold at LHC}},
  \href{https://doi.org/10.1140/epjc/s10052-009-0892-7}{\emph{Eur. Phys. J. C}
  {\bfseries 60} (2009) 375} [\href{https://arxiv.org/abs/0812.0919}{{\ttfamily
  0812.0919}}].

\bibitem{Sumino:2010bv}
Y.~Sumino and H.~Yokoya, \emph{{Bound-state effects on kinematical
  distributions of top quarks at hadron colliders}},
  \href{https://doi.org/10.1007/JHEP09(2010)034}{\emph{JHEP} {\bfseries 09}
  (2010) 034} [\href{https://arxiv.org/abs/1007.0075}{{\ttfamily 1007.0075}}].

\bibitem{Ju:2020otc}
W.-L.~Ju, G.~Wang, X.~Wang, X.~Xu, Y.~Xu and L.L.~Yang, \emph{{Top quark pair
  production near threshold: single/double distributions and mass
  determination}}, \href{https://doi.org/10.1007/JHEP06(2020)158}{\emph{JHEP}
  {\bfseries 06} (2020) 158}
  [\href{https://arxiv.org/abs/2004.03088}{{\ttfamily 2004.03088}}].

\bibitem{Fuks:2021xje}
B.~Fuks, K.~Hagiwara, K.~Ma and Y.-J.~Zheng, \emph{{Signatures of toponium
  formation in LHC run~2 data}},
  \href{https://doi.org/10.1103/PhysRevD.104.034023}{\emph{Phys. Rev. D}
  {\bfseries 104} (2021) 034023}
  [\href{https://arxiv.org/abs/2102.11281}{{\ttfamily 2102.11281}}].

\bibitem{Djouadi:2024lyv}
A.~Djouadi, J.~Ellis and J.~Quevillon, \emph{{Discriminating between
  Pseudoscalar Higgs and Toponium States at the LHC and Beyond}},
  [\href{https://arxiv.org/abs/2412.15138}{{\ttfamily 2412.15138}}].

\bibitem{Moretti:2012mq}
S.~Moretti and D.A.~Ross, \emph{{On the top-antitop invariant mass spectrum at
  the LHC from a Higgs boson signal perspective}},
  \href{https://doi.org/10.1016/j.physletb.2012.04.074}{\emph{Phys. Lett. B}
  {\bfseries 712} (2012) 245}
  [\href{https://arxiv.org/abs/1203.3746}{{\ttfamily 1203.3746}}].

\bibitem{Ellis:1975ap}
J.R.~Ellis, M.K.~Gaillard and D.V.~Nanopoulos, \emph{{A Phenomenological
  Profile of the Higgs Boson}},
  \href{https://doi.org/10.1016/0550-3213(76)90382-5}{\emph{Nucl. Phys. B}
  {\bfseries 106} (1976) 292}.

\bibitem{Cahn:1983ip}
R.N.~Cahn and S.~Dawson, \emph{{Production of Very Massive Higgs Bosons}},
  \href{https://doi.org/10.1016/0370-2693(84)91180-8}{\emph{Phys. Lett. B}
  {\bfseries 136} (1984) 196}.

\bibitem{Spira:1995rr}
M.~Spira, A.~Djouadi, D.~Graudenz and P.M.~Zerwas, \emph{{Higgs boson
  production at the LHC}},
  \href{https://doi.org/10.1016/0550-3213(95)00379-7}{\emph{Nucl. Phys. B}
  {\bfseries 453} (1995) 17}
  [\href{https://arxiv.org/abs/hep-ph/9504378}{{\ttfamily hep-ph/9504378}}].

\bibitem{Plehn:1996wb}
T.~Plehn, M.~Spira and P.M.~Zerwas, \emph{{Pair production of neutral Higgs
  particles in gluon-gluon collisions}},
  \href{https://doi.org/10.1016/0550-3213(96)00418-X}{\emph{Nucl. Phys. B}
  {\bfseries 479} (1996) 46}
  [\href{https://arxiv.org/abs/hep-ph/9603205}{{\ttfamily hep-ph/9603205}}].

\bibitem{Djouadi:2005gi}
A.~Djouadi, \emph{{The Anatomy of electro-weak symmetry breaking. I: The Higgs
  boson in the standard model}},
  \href{https://doi.org/10.1016/j.physrep.2007.10.004}{\emph{Phys. Rept.}
  {\bfseries 457} (2008) 1}
  [\href{https://arxiv.org/abs/hep-ph/0503172}{{\ttfamily hep-ph/0503172}}].

\bibitem{Djouadi:1991tka}
A.~Djouadi, M.~Spira and P.M.~Zerwas, \emph{{Production of Higgs bosons in
  proton colliders: QCD corrections}},
  \href{https://doi.org/10.1016/0370-2693(91)90375-Z}{\emph{Phys. Lett. B}
  {\bfseries 264} (1991) 440}.

\bibitem{Dawson:1990zj}
S.~Dawson, \emph{{Radiative corrections to Higgs boson production}},
  \href{https://doi.org/10.1016/0550-3213(91)90061-2}{\emph{Nucl. Phys. B}
  {\bfseries 359} (1991) 283}.

\bibitem{Kauffman:1993wv}
R.P.~Kauffman and W.~Schaffer, \emph{{Effective Lagrangian for gluons and Higgs
  pseudoscalars}},  in \emph{{Workshop on Physics at Current Accelerators and
  the Supercollider}}, 6, 1993.

\bibitem{Demartin:2014fia}
F.~Demartin, F.~Maltoni, K.~Mawatari, B.~Page and M.~Zaro, \emph{{Higgs
  characterisation at NLO in QCD: CP properties of the top-quark Yukawa
  interaction}},
  \href{https://doi.org/10.1140/epjc/s10052-014-3065-2}{\emph{Eur. Phys. J. C}
  {\bfseries 74} (2014) 3065}
  [\href{https://arxiv.org/abs/1407.5089}{{\ttfamily 1407.5089}}].

\bibitem{Catani:2019hip}
S.~Catani, S.~Devoto, M.~Grazzini, S.~Kallweit and J.~Mazzitelli,
  \emph{{Top-quark pair production at the LHC: Fully differential QCD
  predictions at NNLO}},
  \href{https://doi.org/10.1007/JHEP07(2019)100}{\emph{JHEP} {\bfseries 07}
  (2019) 100} [\href{https://arxiv.org/abs/1906.06535}{{\ttfamily
  1906.06535}}].

\bibitem{Kidonakis:2022hfa}
N.~Kidonakis, \emph{{Higher-order corrections for $t{\bar t}$ production at
  high energies}},  in \emph{{Snowmass 2021}}, 3, 2022
  [\href{https://arxiv.org/abs/2203.03698}{{\ttfamily 2203.03698}}].

\bibitem{Bahl:2022igd}
H.~Bahl, T.~Biek\"otter, S.~Heinemeyer, C.~Li, S.~Paasch, G.~Weiglein et~al.,
  \emph{{HiggsTools: BSM scalar phenomenology with new versions of HiggsBounds
  and HiggsSignals}},  [\href{https://arxiv.org/abs/2210.09332}{{\ttfamily
  2210.09332}}].

\bibitem{Harlander:2012pb}
R.V.~Harlander, S.~Liebler and H.~Mantler, \emph{{SusHi: A program for the
  calculation of Higgs production in gluon fusion and bottom-quark annihilation
  in the Standard Model and the MSSM}},
  \href{https://doi.org/10.1016/j.cpc.2013.02.006}{\emph{Comput. Phys. Commun.}
  {\bfseries 184} (2013) 1605}
  [\href{https://arxiv.org/abs/1212.3249}{{\ttfamily 1212.3249}}].

\bibitem{Harlander:2016hcx}
R.V.~Harlander, S.~Liebler and H.~Mantler, \emph{{SusHi Bento: Beyond NNLO and
  the heavy-top limit}},
  \href{https://doi.org/10.1016/j.cpc.2016.10.015}{\emph{Comput. Phys. Commun.}
  {\bfseries 212} (2017) 239}
  [\href{https://arxiv.org/abs/1605.03190}{{\ttfamily 1605.03190}}].

\bibitem{LHCHiggsCrossSectionWorkingGroup:2016ypw}
{\scshape LHC Higgs Cross Section Working Group} collaboration, \emph{{Handbook
  of LHC Higgs Cross Sections: 4. Deciphering the Nature of the Higgs Sector}},
   [\href{https://arxiv.org/abs/1610.07922}{{\ttfamily 1610.07922}}].

\bibitem{Banfi:2023udd}
A.~Banfi, N.~Kauer, A.~Lind, J.M.~Lindert and R.~Wood, \emph{{Higgs
  interference effects in top-quark pair production in the 1HSM}},
  [\href{https://arxiv.org/abs/2309.16759}{{\ttfamily 2309.16759}}].

\bibitem{Dabelstein:1995js}
A.~Dabelstein, \emph{{Fermionic decays of neutral MSSM Higgs bosons at the one
  loop level}}, \href{https://doi.org/10.1016/0550-3213(95)00523-2}{\emph{Nucl.
  Phys. B} {\bfseries 456} (1995) 25}
  [\href{https://arxiv.org/abs/hep-ph/9503443}{{\ttfamily hep-ph/9503443}}].

\bibitem{Heinemeyer:2000fa}
S.~Heinemeyer, W.~Hollik and G.~Weiglein, \emph{{Decay widths of the neutral CP
  even MSSM Higgs bosons in the Feynman diagrammatic approach}},
  \href{https://doi.org/10.1007/s100520050010}{\emph{Eur. Phys. J. C}
  {\bfseries 16} (2000) 139}
  [\href{https://arxiv.org/abs/hep-ph/0003022}{{\ttfamily hep-ph/0003022}}].

\bibitem{Heinemeyer:2001iy}
S.~Heinemeyer, W.~Hollik, J.~Rosiek and G.~Weiglein, \emph{{Neutral MSSM Higgs
  boson production at e+ e- colliders in the Feynman diagrammatic approach}},
  \href{https://doi.org/10.1007/s100520100631}{\emph{Eur. Phys. J. C}
  {\bfseries 19} (2001) 535}
  [\href{https://arxiv.org/abs/hep-ph/0102081}{{\ttfamily hep-ph/0102081}}].

\bibitem{Denner:1991kt}
A.~Denner, \emph{{Techniques for calculation of electroweak radiative
  corrections at the one loop level and results for W physics at LEP-200}},
  \href{https://doi.org/10.1002/prop.2190410402}{\emph{Fortsch. Phys.}
  {\bfseries 41} (1993) 307} [\href{https://arxiv.org/abs/0709.1075}{{\ttfamily
  0709.1075}}].

\bibitem{Sakurai:2022cki}
K.~Sakurai and W.~Yin, \emph{{Suppression of Higgs mixing by
  \textquotedblleft{}quantum Zeno effect\textquotedblright{}}},
  \href{https://doi.org/10.1140/epjc/s10052-023-11664-y}{\emph{Eur. Phys. J. C}
  {\bfseries 83} (2023) 498}
  [\href{https://arxiv.org/abs/2204.01739}{{\ttfamily 2204.01739}}].

\bibitem{LoChiatto:2024guj}
P.~Lo~Chiatto and F.~Yu, \emph{{Consistent Electroweak Phenomenology of a
  Nearly Degenerate $Z'$ Boson}},
  [\href{https://arxiv.org/abs/2405.03396}{{\ttfamily 2405.03396}}].

\bibitem{basler2020di}
P.~Basler, S.~Dawson, C.~Englert and M.~M\"uhlleitner, \emph{{Di-Higgs boson
  peaks and top valleys: Interference effects in Higgs sector extensions}},
  \href{https://doi.org/10.1103/PhysRevD.101.015019}{\emph{Phys. Rev. D}
  {\bfseries 101} (2020) 015019}
  [\href{https://arxiv.org/abs/1909.09987}{{\ttfamily 1909.09987}}].

\bibitem{Lavoura:1994fv}
L.~Lavoura and J.P.~Silva, \emph{{Fundamental CP violating quantities in a
  SU(2)$\times$U(1) model with many Higgs doublets}},
  \href{https://doi.org/10.1103/PhysRevD.50.4619}{\emph{Phys. Rev. D}
  {\bfseries 50} (1994) 4619}
  [\href{https://arxiv.org/abs/hep-ph/9404276}{{\ttfamily hep-ph/9404276}}].

\bibitem{Botella:1994cs}
F.J.~Botella and J.P.~Silva, \emph{{Jarlskog - like invariants for theories
  with scalars and fermions}},
  \href{https://doi.org/10.1103/PhysRevD.51.3870}{\emph{Phys. Rev. D}
  {\bfseries 51} (1995) 3870}
  [\href{https://arxiv.org/abs/hep-ph/9411288}{{\ttfamily hep-ph/9411288}}].

\bibitem{ElKaffas:2007rq}
A.W.~El~Kaffas, P.~Osland and O.M.~Ogreid, \emph{{CP violation, stability and
  unitarity of the two Higgs doublet model}}, {\emph{Nonlin. Phenom. Complex
  Syst.} {\bfseries 10} (2007) 347}
  [\href{https://arxiv.org/abs/hep-ph/0702097}{{\ttfamily hep-ph/0702097}}].

\bibitem{ATLAS:2020ior}
{\scshape ATLAS} collaboration, \emph{{$\mathcal{CP}$ Properties of Higgs Boson
  Interactions with Top Quarks in the $t\bar{t}H$ and $tH$ Processes Using $H
  \rightarrow \gamma\gamma$ with the ATLAS Detector}},
  \href{https://doi.org/10.1103/PhysRevLett.125.061802}{\emph{Phys. Rev. Lett.}
  {\bfseries 125} (2020) 061802}
  [\href{https://arxiv.org/abs/2004.04545}{{\ttfamily 2004.04545}}].

\bibitem{CMS:2020cga}
{\scshape CMS} collaboration, \emph{{Measurements of $\mathrm{t\bar{t}}H$
  Production and the CP Structure of the Yukawa Interaction between the Higgs
  Boson and Top Quark in the Diphoton Decay Channel}},
  \href{https://doi.org/10.1103/PhysRevLett.125.061801}{\emph{Phys. Rev. Lett.}
  {\bfseries 125} (2020) 061801}
  [\href{https://arxiv.org/abs/2003.10866}{{\ttfamily 2003.10866}}].

\bibitem{Bahl:2020wee}
H.~Bahl, P.~Bechtle, S.~Heinemeyer, J.~Katzy, T.~Klingl, K.~Peters et~al.,
  \emph{{Indirect $\mathcal{CP}$ probes of the Higgs-top-quark interaction:
  current LHC constraints and future opportunities}},
  \href{https://doi.org/10.1007/JHEP11(2020)127}{\emph{JHEP} {\bfseries 11}
  (2020) 127} [\href{https://arxiv.org/abs/2007.08542}{{\ttfamily
  2007.08542}}].

\bibitem{CMS:2022uox}
{\scshape CMS} collaboration, \emph{{Constraints on anomalous Higgs boson
  couplings to vector bosons and fermions from the production of Higgs bosons
  using the \ensuremath{\tau}\ensuremath{\tau} final state}},
  \href{https://doi.org/10.1103/PhysRevD.108.032013}{\emph{Phys. Rev. D}
  {\bfseries 108} (2023) 032013}
  [\href{https://arxiv.org/abs/2205.05120}{{\ttfamily 2205.05120}}].

\bibitem{CMS:2022dbt}
{\scshape CMS} collaboration, \emph{{Search for $CP$ violation in ttH and tH
  production in multilepton channels in proton-proton collisions at $\sqrt{s}$
  = 13 TeV}}, \href{https://doi.org/10.1007/JHEP07(2023)092}{\emph{JHEP}
  {\bfseries 07} (2023) 092}
  [\href{https://arxiv.org/abs/2208.02686}{{\ttfamily 2208.02686}}].

\bibitem{Bahl:2022yrs}
H.~Bahl, E.~Fuchs, S.~Heinemeyer, J.~Katzy, M.~Menen, K.~Peters et~al.,
  \emph{{Constraining the ${\mathcal {C}}{\mathcal {P}}$ structure of
  Higgs-fermion couplings with a global LHC fit, the electron EDM and
  baryogenesis}},
  \href{https://doi.org/10.1140/epjc/s10052-022-10528-1}{\emph{Eur. Phys. J. C}
  {\bfseries 82} (2022) 604}
  [\href{https://arxiv.org/abs/2202.11753}{{\ttfamily 2202.11753}}].

\bibitem{Brod:2022bww}
J.~Brod, J.M.~Cornell, D.~Skodras and E.~Stamou, \emph{{Global constraints on
  Yukawa operators in the standard model effective theory}},
  \href{https://doi.org/10.1007/JHEP08(2022)294}{\emph{JHEP} {\bfseries 08}
  (2022) 294} [\href{https://arxiv.org/abs/2203.03736}{{\ttfamily
  2203.03736}}].

\bibitem{Bahl:2023qwk}
H.~Bahl, E.~Fuchs, M.~Hannig and M.~Menen, \emph{{Classifying the CP properties
  of the ggH coupling in H+2j production}},
  [\href{https://arxiv.org/abs/2309.03146}{{\ttfamily 2309.03146}}].

\bibitem{ATLAS:2023cbt}
{\scshape ATLAS} collaboration, \emph{{Probing the CP nature of the
  top\textendash{}Higgs Yukawa coupling in $t\overline{t}H$ and $tH$ events
  with H\textrightarrow{} $b\overline{b}$ decays using the ATLAS detector at
  the LHC}}, \href{https://doi.org/10.1016/j.physletb.2024.138469}{\emph{Phys.
  Lett. B} {\bfseries 849} (2024) 138469}
  [\href{https://arxiv.org/abs/2303.05974}{{\ttfamily 2303.05974}}].

\bibitem{Alwall:2014hca}
J.~Alwall, R.~Frederix, S.~Frixione, V.~Hirschi, F.~Maltoni, O.~Mattelaer
  et~al., \emph{{The automated computation of tree-level and next-to-leading
  order differential cross sections, and their matching to parton shower
  simulations}}, \href{https://doi.org/10.1007/JHEP07(2014)079}{\emph{JHEP}
  {\bfseries 07} (2014) 079} [\href{https://arxiv.org/abs/1405.0301}{{\ttfamily
  1405.0301}}].

\bibitem{Ball:2012cx}
R.D.~Ball et~al., \emph{{Parton distributions with LHC data}},
  \href{https://doi.org/10.1016/j.nuclphysb.2012.10.003}{\emph{Nucl. Phys. B}
  {\bfseries 867} (2013) 244}
  [\href{https://arxiv.org/abs/1207.1303}{{\ttfamily 1207.1303}}].

\bibitem{Buckley:2014ana}
A.~Buckley, J.~Ferrando, S.~Lloyd, K.~Nordstr\"om, B.~Page, M.~R\"ufenacht
  et~al., \emph{{LHAPDF6: parton density access in the LHC precision era}},
  \href{https://doi.org/10.1140/epjc/s10052-015-3318-8}{\emph{Eur. Phys. J. C}
  {\bfseries 75} (2015) 132} [\href{https://arxiv.org/abs/1412.7420}{{\ttfamily
  1412.7420}}].

\bibitem{mg-ggH-ggA-loop}
D.B.~Franzosi and C.~Zhang, \emph{Bottom and top loop structure in ggh and
  gga},  2014.
\newblock
  \url{https://cp3.irmp.ucl.ac.be/projects/madgraph/wiki/Models/ggHFullLoop#no1}.

\end{thebibliography}\endgroup
\end{document}